\documentclass[twocolumn,epjc3]{svjour3}  
\smartqed
\usepackage[utf8,latin1]{inputenc}
\newcommand{\qm}[1]{``#1''}
\usepackage{amsmath}
\usepackage{lipsum}
\usepackage{ulem}
\usepackage{microtype}
\usepackage{multicol}
\usepackage{amssymb}
\usepackage{mathabx}
\usepackage{mathrsfs}  
\usepackage{cases}
\usepackage{revsymb}
\usepackage{tabularx}
\usepackage{colortbl}
\usepackage{graphicx}
\usepackage{dcolumn}
\usepackage{xcolor}
\usepackage{bm}
\usepackage{widetext}
\usepackage{cuted}
\usepackage{hyperref}
\hypersetup{colorlinks, linkcolor={red},citecolor={blue},urlcolor={blue}}  
\usepackage[square,numbers,sort&compress]{natbib}
\usepackage{hyperref}

\newcommand{\dd}{{\rm d}}

\newcommand{\OC}[1]{{\rm O}\left(c^{#1}\right)}

\journalname{Eur. Phys. J. C}

\begin{document}

\title{Radiative losses and  radiation-reaction effects at the first post-Newtonian order in Einstein-Cartan theory}
\titlerunning{Radiative losses and  radiation-reaction effects at the first post-Newtonian order in Einstein-Cartan theory}

\author{Vittorio De Falco\thanksref{e1,addr1,addr2}
\and
Emmanuele Battista\thanksref{e2,e3,addr3}
\and
Davide Usseglio\thanksref{e4,addr1,addr2}
\and
Salvatore Capozziello\thanksref{e5,addr1,addr2,addr4}}

\thankstext{e1}{e-mail: vittorio.defalco-ssm@unina.it}
\thankstext{e2}{e-mail: emmanuele.battista@univie.ac.at}
\thankstext{e3}{e-mail: emmanuelebattista@gmail.com}
\thankstext{e4}{e-mail: davide.usseglio-ssm@unina.it}
\thankstext{e5}{e-mail: capozziello@na.infn.it}

\authorrunning{De Falco et al. (2023)}

\institute{Scuola Superiore Meridionale, Largo San Marcellino 10, 80138 Napoli, Italy\label{addr1}
\and
Istituto Nazionale di Fisica Nucleare, Sezione di Napoli, Complesso Universitario di Monte S. Angelo, Via Cintia Edificio 6, 80126 Napoli, Italy \label{addr2}
\and
Department of Physics, University of Vienna, Boltzmanngasse 5, A-1090 Vienna, Austria \label{addr3}
\and
Dipartimento di Fisica \qm{E. Pancini}, Universit\'a di Napoli \qm{Federico II}, Complesso Universitario di Monte S. Angelo, Via Cintia Edificio 6, I-80126 Napoli, Italy \label{addr4}}

\date{Received: \today / Accepted: }

\maketitle

\begin{abstract}
Gravitational radiation-reaction phenomena occurring in the dynamics of inspiralling compact binary systems are investigated  at the first post-Newtonian order beyond the quadrupole approximation in the context of Einstein-Cartan theory, where quantum spin effects are modeled via the Weyssenhoff fluid. We exploit balance equations for the energy and angular momentum to determine the binary orbital decay until the two bodies collide. Our framework deals with both quasi-elliptic and quasi-circular trajectories, which are then smoothly connected. Key observables like the laws of variation of the orbital phase and frequency characterizing the quasi-circular motion are derived analytically. We conclude our analysis with an estimation of the spin contributions at the merger, which are examined both in the time domain and the Fourier frequency space through the stationary wave approximation. 
\end{abstract}

\normalem
\section{Introduction}
\label{sec:intro}
Gravitational-wave (GW) astronomy represents an essential branch of modern astrophysics. In this context, GWs are employed to inquire gravitational phenomena either within General Relativity (GR) or alternative frameworks, as well as to gather information on compact objects (i.e, neutron stars (NSs) and black holes (BHs)), high-energy events (such as core collapse supernovae), and primordial processes occurring in the early universe shortly after the Big Bang \cite{Sathyaprakash2009,Buonanno2014aza,Mastrogiovanni2022}. Since their first direct detection occurred in September 2015, LIGO and Virgo have observed about one hundred GW events from astrophysical compact binary mergers \cite{Abbott2022}. Currently, LIGO, Virgo, and KAGRA are undergoing an update period to further increase their sensitivity and reduce the signal-to-noise ratio in view of the forthcoming fourth observing run \cite{Cahillane2022}, where the GW detection rate is expected to be four times larger than the previous one \cite{KAGRA2013-review}.   In this unquenchable observational GW campaign, fundamental partners will be the future third-generation ground-based GW interferometers  Einstein Telescope \cite{ET_SCIENCE2020JCAP} and Cosmic Explorer \cite{Reitze2019a,Reitze2019b}, and the space-based devices LISA \cite{Audley2017} and TianQin \cite{TianQin2015}.

The existence of GWs was predicted for the very first time in 1916 by Einstein, who also worked out the famous quadrupole formula two years later \cite{Poisson-Will2014}. In high-energy astrophysics, GWs are generated by inspiralling compact binaries (classified in BH--BH, NS--NS, BH--NS systems), whose dynamics is influenced by the dissipative effect of gravitational radiation-reaction forces. These damping terms are responsible for the secular loss of energy, momentum, and angular momentum of the system. The net result of this irreversible process is that the body trajectories tend to circularize very rapidly, with a consequent decrease of their separation and increase of the orbital frequency \cite{Maggiore:GWs_Vol1,Poisson-Will2014}. Radiation-reaction contributions are odd under the time-reversal operation and arise in the equations of motion of a self-gravitating system at five halves post-Newtonian (2.5PN) approximation in GR \cite{Maggiore:GWs_Vol1}. At this level, they stem from the well-known Burke-Thorne potential, which involves the fifth-order time derivative of the (Newtonian) quadrupole moment of the source \cite{Burke1969,Thorne1969}. As shown in Ref. \cite{Blanchet1993}, the occurrence of gravitational back-reaction effects inside the source can be inferred from the analysis of the antisymmetric component of the exterior gravitational field, namely a retarded-minus-advanced solution of the wave equation (see Ref. \cite{Blanchet2014}, for further details).

The first observational confirmation of the existence of gravitational dissipative forces triggered by the emission of GWs came from the monitoring of the Hulse-Taylor binary pulsar in the mid seventies \cite{Hulse1975,Taylor1979}. This event shows that the behaviour of Hulse-Taylor-like systems can be reliably accounted for by considering the approach relying on the Burke-Thorne force, whereas for relativistic inspiralling compact binaries higher post-Newtonian (PN) approximations are necessary for constructing more accurate theoretical templates \cite{Blanchet1996}. \emph{The addition of the $n$th PN contributions to the radiation-reaction forces corresponds to the ($n+2.5$)th PN corrections in the equations of motion} \cite{Blanchet1995,Blanchet1997}. The complexity in attaining high PN orders can be circumvented via the \emph{balance equations}, which relate the losses of energy, angular momentum, and linear momentum of a system to the corresponding fluxes in the radiation field in the wave zone of the source \cite{Blanchet1996,Maggiore:GWs_Vol1,Blanchet2014}. 

The secular evolution of PN isolated sources has been first described in the pioneering works of Peters and Mathews via balance arguments \cite{Peters1963,Peters1964}. Such a method yields in general enormous advantages, as it permits to evaluate the $n$th PN correction to the radiation-reaction forces (or, equivalently, the $n$th PN correction to the quadrupole approximation) even without controlling the corresponding equations of motion of the system with $(n+2.5)$PN accuracy \cite{Blanchet-Schafer1989,Blanchet1995,Blanchet1997}. The correctness of this  heuristic scheme  for the case $n=0$ has been proved by Damour by employing the explicit form of the 2.5PN equations of motion \cite{Damour1983}. The extension up to the order with $n=1.5$ has been then provided by Blanchet, who has demonstrated that \emph{flux-balance equations for energy, angular momentum, linear momentum, and center-of-mass position are mathematically consistent in GR at 1.5PN order in the radiation field} \cite{Blanchet1997,Bernard2017,Blanchet-Faye2019,Blanchet2023a,Blanchet2023b}. At this level, the underlying radiation-reaction force involves the contribution of the GW tails, which are a manifestation of the \emph{nonlinear character} of GR \cite{Blanchet1988,Blanchet1993,Maggiore:GWs_Vol1,Blanchet2014}.

The PN technique and the point-like approximation can characterize  the decay of binary systems only during the inspiral phase. In fact, these methodologies break down in the merger and ringdown stages, where the tidal forces and the internal structure of the bodies become fundamental. In addition, PN expansions formally do not converge near the light ring, due to the high velocities attained by the bodies. In this scenario, one must resort to the effective-one-body (EOB) framework  \cite{Buonanno1999,Buonanno2000,Damour2011}. However, the complicated mathematical structure of this model makes it convenient to employ the PN and the point-particle schemes all the way down to the coalescence, in order to infer a rough estimate of the effects under consideration. This approach can be especially useful in the context of extended theories of gravity, where, to the best of our knowledge, the back-reaction phenomenon has been treated only in linearized $f(R)$ gravity \cite{DeLaurentis2011,Narang2022} and in scalar tensor models with 1.5PN accuracy  \cite{Bernard2022}.

Our research activity is framed in Einstein-Cartan (EC) theory, a generalization of GR where both mass and quantum spin of matter are the source of the gravitational field \cite{Hehl1976_fundations}. In this context, we have developed a structured program, where we have investigated at 1PN level: the GW generation problem by means of the Blanchet-Damour formalism \cite{Paper1} with an explicit application to the Weyssenhoff semiclassical model of a neutral spinning perfect fluid \cite{Paper2}; the $N$-body translational and rotational dynamics by applying the point-particle limit to the continuous description provided by  the Weyssenhoff fluid \cite{Paper3,Paper4}; finally, we have derived the analytical expressions of the relative orbit and coordinate time pertaining to the dynamics of binary systems having aligned spins together with some applications \cite{Paper5}.

In this paper, our primary goal is to determine the dynamics of a binary system influenced by back-reaction forces through the above mentioned balance scheme with $n=1$ PN accuracy. In our analysis, the bodies are endowed with a quantum spin modelled via the Weyssenhoff fluid. Our results are valid for spinning, weakly self-gravitating, slowly moving, and weakly stressed sources (i.e., PN sources in EC theory). However, our framework can be applied also to compact binary systems, since we have found that the effacing  principle   is valid, at least at 1PN level, also in EC model \cite{Paper3,Paper4}. 

The manuscript is organized as follows. We  recall  some concepts and preliminary topics  in Sec. \ref{sec:math_preliminaries}. Then, in Sec. \ref{sec:BR_model} we develop, for the first time in the literature, the back-reaction model pertaining to binary systems at 1PN order in EC theory. Here, we calculate  the decay of the orbital parameters, and elaborate the ensuing  gravitational waveform and  orbit. In Sec. \ref{sec:results}, we exploit our approach to  estimate the order of magnitude of the EC spin effects at the merger. Finally, we draw the conclusions and outline future perspectives in Sec. \ref{sec:conclusions}.

\subsubsection*{Notations and conventions}
Greek indices take values  $0,1,2,3$, while lowercase Latin ones $1,2,3$. Spacetime coordinates are $x^\mu = (ct,\boldsymbol{x})$. Four-vectors are written as $a^\mu = (a^0,\boldsymbol{a})$, and $\boldsymbol{a} \cdot \boldsymbol{b}:= \delta_{lk}a^l b^k$, $\vert \boldsymbol{a} \vert\equiv a :=  \left(\boldsymbol{a} \cdot \boldsymbol{a}\right)^{1/2}$, and $\left(\boldsymbol{a} \times \boldsymbol{b}\right)^i := \varepsilon_{ilk} a^l b^k$, where $\varepsilon_{kli}$ is the total antisymmetric Levi-Civita symbol. The symmetric-trace-free (STF) projection of a tensor $A^{ij\dots k}$ is indicated with $A^{\langle ij\dots k \rangle }$. Superscripts $(l)$ denote $l$ successive time derivatives. $L=i_1 i_2 \dots i_l$ denotes a multi-index consisting of $l$ spatial indices. Round (respectively, square) brackets around a pair of indices stands for the usual symmetrization (respectively, antisymmetrization) procedure, i.e., $A_{(ij)}=\frac{1}{2}(A_{ij}+A_{ji})$ (respectively, $A_{[ij]}=\frac{1}{2}(A_{ij}-A_{ji})$). 

\section{Preliminaries}
\label{sec:math_preliminaries}

This section is devoted to recall some basic concepts from our previous studies, which are fundamental for the comprehension of the paper.
We start with a brief summary of the main features of  EC model (see Sec. \ref{Sec:EC-theory}). Then, we give the general 1PN-accurate expressions of the   gravitational waveform and fluxes  in Sec. \ref{sec:waveform-fluxes}, whereas some useful formulas pertaining to binary systems are presented in Sec. \ref{Sec:binary-systems}.

Hereafter, the  bodies and all the related quantities are labelled with capital Latin indices  $A,B=1,2$.

\subsection{Einstein-Cartan model}
\label{Sec:EC-theory}
EC  theory is defined on a  Riemann-Cartan spacetime endowed with a symmetric metric tensor $g_{\alpha \beta}$ and the most general metric-compatible affine connection $\Gamma^\lambda_{\mu \nu}$ \cite{Hehl1976_fundations,Gasperini-DeSabbata}. Its antisymmetric part  $\Gamma^\lambda_{[\mu \nu]}:= S_{\mu \nu}^{\ \ \ \lambda}$ defines the \emph{Cartan torsion tensor}, which represents the geometrical counterpart of the quantum spin of elementary particles (see Ref. \cite{Capozziello2001}, for an extensive discussion on the geometric and physical aspects of  torsion).  

The matter content of EC field equations (which, we recall, can be written in a form resembling Einstein equations) is described in terms  of the  combined energy-momentum tensor $\Theta^{\alpha\beta}$, which reads as
\begin{align}
\Theta^{\alpha\beta}:=T^{\alpha\beta}+\frac{8 \pi G}{c^4}\mathcal{S}^{\alpha \beta},
\label{Theta-alpha-beta}  
\end{align}
where $T^{\alpha\beta}$ is the metric energy-momentum tensor and we dub  $\mathcal{S}^{\alpha \beta}$ the torsional energy-momentum tensor. 

In our setup, we will suppose that the spin of continuous matter can be characterized by the semiclassical Weyssenhoff model endowed with the Frenkel condition (which leads to the relation $S_{\mu\nu}{}^\nu=0$). This has important consequences for the study of both the GW generation problem and the hydrodynamics in EC theory. For details see Refs. \cite{Paper1,Paper2} and references therein.

\subsection{Gravitational waveform and fluxes}
\label{sec:waveform-fluxes}
The wave zone is the spatial region of $\mathbb{R}^3$ where a detector apparatus is framed \cite{Maggiore:GWs_Vol1,Blanchet2014}. In this domain, one of the main observables is represented by the  \emph{asymptotic waveform} $\mathscr{H}^{\rm TT}_{ij}$, which at 1PN level reads as \cite{Paper1,Paper2}
\begin{align} \label{gravitational_wave_amplitude}
    \mathscr{H}_{ij}^{\rm TT}(x^\mu) & = \dfrac{2G}{c^4 \vert\boldsymbol{x}\vert} \mathscr{P}_{ijkl}(\boldsymbol{n})  \Biggr\{ \overset{(2)}{I}{}_{kl}^{{\rm rad}}(\mathfrak{u}) 
    \nonumber \\
    & + \dfrac{1}{c} \left[  \dfrac{1}{3} n_a  \overset{(3)}{I}{}_{kla}^{{\rm rad}}(\mathfrak{u}) +\dfrac{4}{3} n_b \epsilon_{ab(k} \overset{(2)}{J}{}_{l)a}^{{\rm rad}}(\mathfrak{u})  \right]
    \nonumber \\
    & +\dfrac{1}{c^2} \left[\dfrac{1}{12}n_{a}n_{b}   \overset{(4)}{I}{}_{klab}^{{\rm rad}}(\mathfrak{u}) \right.
    \nonumber \\
     &\left. + \dfrac{1}{2}n_{b}n_{c} \epsilon_{ab(k}   \overset{(3)}{J}{}_{l)ac}^{{\rm rad}}(\mathfrak{u})  \right]
      + {\rm O}(c^{-3}) \Biggr\}, 
\end{align}
where $\mathfrak{u}= t-\vert\boldsymbol{x}\vert/c$,  $\boldsymbol{n}=\boldsymbol{x}/\vert\boldsymbol{x}\vert$, and $\mathscr{P}_{ijkl}(\boldsymbol{n})$ is transverse-traceless (TT) projection operator onto the plane orthogonal to $\boldsymbol{n}$; furthermore, $I^{\rm rad}_L$ and  $J^{\rm rad}_L$ denote the STF mass-type and  current-type radiative multipole moments of order $l$, respectively.

GWs transport  energy and angular momentum away from their source. This phenomenon can be accounted for via the  \textit{total radiated power} $\mathcal{F}$ (also dubbed \textit{energy flux} or \textit{gravitational luminosity}) and the \emph{angular momentum flux} $\mathcal{G}_i$ associated with the waveform \eqref{gravitational_wave_amplitude}.   At 1PN order, the former can be written as \cite{Paper1,Paper2}
\begin{align} \label{power-radiated-1PN}
\mathcal{F}(t) & = \dfrac{G}{c^5} \Biggl\{\dfrac{1}{5}\overset{(3)}{I}{}_{ij}^{{\rm rad}}\overset{(3)}{I}{}^{{\rm rad}}_{ij} + \dfrac{1}{c^2} \Biggl[ \dfrac{1}{189} \overset{(4)}{I}{}^{{\rm rad}}_{ijk} \overset{(4)}{I}{}^{{\rm rad}}_{ijk}
 \nonumber \\
&+ \dfrac{16}{45} \overset{(3)}{J}{}^{\rm rad}_{ij}\overset{(3)}{J}{}_{ij}^{\rm rad} \Biggr]
+ {\rm O}\left(c^{-4}\right) \Biggr \};
\end{align}
having verified that the GR formula \cite{Maggiore:GWs_Vol1,Blanchet2014} of the angular momentum flux is valid also in EC theory; instead the 1PN-accurate expression of $\mathcal{G}_i$ reads as
\begin{align}\label{ang-momentum-flux-1PN}
\mathcal{G}_i(t)   &= \dfrac{G}{c^5} \varepsilon_{ijk} \Biggl\{\dfrac{2}{5} \overset{(2)}{I}{}^{{\rm rad}}_{jl}\overset{(3)}{I}{}^{{\rm rad}}_{kl} 
   + \dfrac{1}{c^2} \Biggl[ \dfrac{1}{63} \overset{(3)}{I}{}^{{\rm rad}}_{jlp} \overset{(4)}{I}{}^{{\rm rad}}_{klp}
 \nonumber \\
&+ \dfrac{32}{45} \overset{(2)}{J}{}^{{\rm rad}}_{jl}\overset{(3)}{J}{}^{{\rm rad}}_{kl} \Biggr]
+ {\rm O}\left(c^{-4}\right) \Biggr \}.
\end{align}

Note that in the above expressions we have employed harmonic coordinates $x^\mu$, since they  do not differ from radiative ones at 1PN level \cite{Blanchet-Damour1989,Blanchet2014}. Moreover, it is worth pointing out that the resolution of the GW generation problem at  1PN order permits to express the radiative multipole moments in terms of the material features of the source, i.e., $I^{\rm rad}_L$ and  $J^{\rm rad}_L$ can be written as well-defined integral formulas extending over the combined stress-energy tensor $\Theta^{\alpha \beta}$ of the source (cf. Eq. \eqref{Theta-alpha-beta} and see Sec. III G in Ref. \cite{Paper1}, for further details).  

\subsection{Binary systems} 
\label{Sec:binary-systems}
Compact binary systems are the main source of GWs in high-energy astrophysics. As explained in Refs. \cite{Paper2,Paper3,Paper4}, they can be modelled by applying the point-particle limit to the underlying fluid description provided by the Weyssenhoff model.

The 1PN-accurate expressions of the radiative multipole moments, which permit to evaluate the waveform and fluxes presented in Sec. \ref{sec:waveform-fluxes}, can be easily written in  a mass-centered coordinate system. Here,  the motion of each body is constructed via the position vectors \cite{Paper2}
\begin{subequations}
\label{position-vectors-r1-r2-with-spin}
\begin{align}
\boldsymbol{r}_1(t)&=\left[\frac{\mu}{m_1}+\frac{\mu (m_1-m_2)}{2M^2c^2}\left(V^2-\frac{GM}{R}\right)\right]\boldsymbol{R}(t)
\notag\\
&+\frac{2 \nu}{c^2}\left[\dfrac{\boldsymbol{s}_1(t)}{m_1} -\dfrac{\boldsymbol{s}_2(t)}{m_2}\right]\times \boldsymbol{V}(t)+{\rm O}\left(c^{-4}\right),
\\
\boldsymbol{r}_2(t)&=\left[-\frac{\mu}{m_2}+\frac{\mu (m_1-m_2)}{2M^2c^2}\left(V^2-\frac{GM}{R}\right)\right]\boldsymbol{R}(t)\notag\\
&+\frac{2 \nu}{c^2}\left[\dfrac{\boldsymbol{s}_1(t)}{m_1} -\dfrac{\boldsymbol{s}_2(t)}{m_2}\right]\times \boldsymbol{V}(t)+{\rm O}\left(c^{-4}\right),
\end{align}
\end{subequations}
which  yield   \cite{Paper2,Paper5} 
\begin{align}
I^{\rm rad}_{ij}& =\mu R_{\langle ij\rangle }\left[1+\frac{29}{42c^2}(1-3\nu)V^2-\frac{(5-8\nu)}{7c^2}\frac{GM}{R}\right]\notag\\
&+\frac{\mu(1-3\nu)}{21c^2}\left[11R^2 V_{\langle ij\rangle } -12 (\boldsymbol{R} \cdot \boldsymbol{V}) R_{\langle i} V_{j \rangle }\right]\notag\\
&+\frac{8\nu}{3c^2}\left[2(\boldsymbol{V}\times\boldsymbol{\sigma})^{\langle i}R^{j\rangle }  -\left( \boldsymbol{R}\times\boldsymbol{\sigma}\right)^{\langle i}V^{j \rangle }\right]\notag\\
&+{\rm O}\left(c^{-3}\right),
\label{I-ij-rad-pp-limit-expression-2}\\
I^{\rm rad}_{ijk}& =-\mu\sqrt{1-4\nu}R_{\langle ijk\rangle }+{\rm O}\left(c^{-2}\right),
\\
I^{\rm rad}_{ijkl}& =\mu(1-3\nu)R_{\langle ijkl\rangle} +{\rm O}\left(c^{-2}\right),
\\
J^{\rm rad}_{ij}& =-\mu\sqrt{1-4\nu}\epsilon_{kl\langle i}R_{j\rangle k}V_l\notag\\
&+3\mu\left(\frac{s_1^{\langle i}R^{j\rangle }}{m_1}-\frac{s_2^{\langle i}R^{j\rangle }}{m_2}\right)
+{\rm O}\left(c^{-2}\right),
\label{J-ij-rad-pp-limit-expression-2}
\\
J^{\rm rad}_{ijk}&=\mu(1-3\nu)R_{\langle ij}\epsilon_{k\rangle lp}R_l V_p +4\nu  R^{\langle i} R^j \sigma^{k \rangle}+{\rm O}\left(c^{-2}\right).
\label{J-ijk-rad-pp-limit-expression-2}
\end{align}
In the above equations,  we have defined the relative vectors  (in   harmonic coordinates)
\begin{align}
\boldsymbol{R}:=\boldsymbol{r}_1-\boldsymbol{r}_2,  \qquad \boldsymbol{V}:= \dfrac{\dd}{\dd t}\boldsymbol{R} = \boldsymbol{v}_1 - \boldsymbol{v}_2,
\end{align}
where  $\boldsymbol{v}_A$  is the velocity vector of the bodies; furthermore, $m_A$ and $\boldsymbol{s}_A$  are the (conserved)  mass(-energy) and spin of the $A$-th object; in addition, 
\begin{align}
    M := m_1 + m_2, \qquad \mu := \dfrac{m_1 m_2}{M}, \qquad \nu := \dfrac{\mu}{M},
\end{align}
are total mass, reduced mass, and symmetric mass ratio of the system, respectively; lastly, we have defined 
\begin{align}
\boldsymbol{\sigma}:= \dfrac{m_2}{m_1}\boldsymbol{s}_1 + \dfrac{m_1}{m_2}\boldsymbol{s}_2.
\end{align}

\section{Back-reaction model in Einstein-Cartan theory: a heuristic approach}
\label{sec:BR_model}
In this section, we develop the back-reaction\footnote{The back-reaction of GWs affects the motion of the source, inducing the inspiral and the coalescence of binary systems. In this process, the emission and propagation of GWs themselves are also affected. From a field-theoretical point of view, this entails the analysis of back-scattering of gravitons on the background spacetime, graviton-graviton scattering, etc (see Ref. \cite{Maggiore:GWs_Vol1}, for more details). } model in EC theory at 1PN order via the heuristic approach. This represents the original part of our work.

In our setup, we will assume that the spins of the bodies are aligned perpendicular to the orbital plane $x-y$, viz. they are directed along the $z$-axis. In this configuration, there is no spin precession, as the rotational dynamics is ruled by the equation $\dd \bar{\boldsymbol{s}}_A /\dd t = \OC{-4}$, where $\bar{\boldsymbol{s}}_A = \boldsymbol{s}_A + \OC{-2}$ is the refined spin which keeps its magnitude constant during the motion (see Refs. \cite{Paper4,Paper5}, for further details). This is a reasonable assumption, because a significant spin precession would take a time span much larger than the coalescence timescale. In our hypotheses, we can thus write $\boldsymbol{s}_A=(0,0,s_{Az})$ and introduce  the following useful definitions, which will facilitate the subsequent  calculations: 
\begin{subequations}
\begin{align}
s_z&:=s_{1z}+s_{2z}, \\
\xi_z&:=\frac{m_2s_{1z}+m_1s_{2z}}{\sqrt{m_1m_2}},
\end{align}    
\end{subequations}
where the spin of each body is assigned as follows \cite{Paper1,Paper2}
\begin{align}
\label{szi-components-BH-NS}
s_{Az}=\begin{cases}  \mathcal{N}\hbar\frac{4\pi}{3}\left(\frac{6Gm_A}{c^2}\right)^3, &{\rm for\ NSs},\\
\mathcal{N}\hbar\frac{4\pi}{3}\left(\frac{2Gm_A}{c^2}\right)^3, &{\rm for\ BHs},
\end{cases}
\end{align}
 $\mathcal{N}=10^{44}\ \mbox{m}^{-3}$ being estimated as the inverse of the nucleon volume \cite{Paper1,Paper2}. 

The first step of our framework consists in  parametrizing the two-body quasi-elliptic motion in terms of two PN orbital parameters (generally represented by the semi-major axis and  eccentricity). The second stage reckons on the calculation of the source energy $E_{\rm source}:=\mu E$ and orbital angular momentum $L_{\rm source} := \mu L$, together with the time averages\footnote{The time average of a function $f(t)$ over a period $P$ is defined as $\langle f(t)\rangle=\frac{1}{P}\int_0^{P}f(t)\dd t$.} $\left<\mathcal{F}\right>$ and $\left<\mathcal{G}\right>$ of the energy and angular momentum fluxes, respectively  (cf. Eqs. \eqref{power-radiated-1PN} and  \eqref{ang-momentum-flux-1PN}). In this way, we are able to construct the balance equations 
\begin{subequations}\label{balance-equations}
\begin{align}
  \frac{\dd E_{\rm source}}{\dd t}&=-\left<\mathcal{F}\right>, 
  \label{E_balance}
  \\
 \frac{\dd L_{\rm source}}{\dd t}&=-\left<\mathcal{G}\right>,
 \label{J_balance}
\end{align}    
\end{subequations}
relating the source energy and orbital angular momentum loss rates to the corresponding averaged fluxes of radiation in the wave zone of the system. The above relations yield two coupled ordinary differential equations (one for the semi-major axis and the other for the eccentricity) and govern the decay evolution of the binary system, as we will explain in Sec. \ref{sec:BR_EC_1PN_elliptic}. When the orbit circularizes (i.e, the eccentricities become zero), we smoothly connect the quasi-elliptic motion with the quasi-circular one. In the latter case, we end up with only one  ordinary differential equation, which inflects the time variation of the orbital radius. This type of dynamics continues up to the coalescence of the two bodies (see Sec. \ref{sec:BR_EC_1PN_circular}). To show how our abovementioned procedure works, we provide an application to an astrophysical GW event in Sec. \ref{sec:model_application}.

The input variables of our model are 
\begin{equation} \label{eq:model_parameters}
\left\{m_1,m_2,a_0,e_0,\theta_0,t_0,d\right\},    
\end{equation}
where $a_0$, $e_0$, $\theta_0$, and $t_0$ are the  initial orbital separation, eccentricity,    polar angle, and  time, respectively, while $d:= \vert \boldsymbol{x} \vert $ denotes  the distance between the GW detector and the gravitational source. Furthermore, the outputs are represented by the  orbital parameters' evolution and the underlying gravitational waveform, as well as the values of the circularization and coalescence times. 

\subsection{Quasi-elliptic orbit}
\label{sec:BR_EC_1PN_elliptic}

In this section, we deal with the quasi-elliptic motion at 1PN order. We first parametrize the orbit \emph{\'a la} Damour-Deruelle (see Sec. \ref{sec:DD_parametrization}). Then the source energy and angular momentum along with the time averages of the radiative fluxes are derived in Secs. \ref{sec:source_E_and_J} and \ref{sec:mean_fluxes}, respectively. In this way, we can write the balance relations (see below Eq. \eqref{balance-equations}) from which we extract the decay equations for the semi-major axis and eccentricity (see Sec. \ref{sec:balance_equations}). After having solved these numerically, we finally compute the two-body orbits and the emitted waveform in Sec. \ref{sec:orbit_and_waveform}.

\subsubsection{Damour-Deruelle parametrization}
\label{sec:DD_parametrization}

To develop our back-reaction model, it is essential to first provide a suitable  parametrization of the 1PN binary motion. 

In our hypotheses, the two objects experience no spin precession, since the total specific angular momentum $\boldsymbol{J}:=\boldsymbol{L}+\mu^{-1}(\bar{\boldsymbol{s}}_1+\bar{\boldsymbol{s}}_2)$ is conserved, at 1PN order, because it is the sum of the two first integrals $\boldsymbol{L}$ and $\bar{\boldsymbol{s}}_1+\bar{\boldsymbol{s}}_2$. For this reason, the two-body translational dynamics can be written starting from the constants of motion $E$ and $L$ (i.e., the specific energy and orbital angular momentum, respectively; see Eqs. (45) and (48) in Ref. \cite{Paper4} and Sec. III A in Ref. \cite{Paper5}, for details). In this way, we can follow the procedure devised by Damour and Deruelle in  Ref. \cite{Damour1985}. Thus, upon  performing the ensuing calculations in polar coordinates $(R,\theta)$, we find that the 1PN relative dynamics in EC theory is ruled by \cite{Paper5} 
\begin{subequations} \label{eqs_EC_split}
\begin{align}
\left(\frac{\dd R}{\dd t}\right)^2&=\mathcal{A}+\frac{2\mathcal{B}}{R}+\frac{\mathcal{C}}{R^2}+\frac{\mathcal{D}}{R^3}+\OC{-4},
\\
\frac{\dd \theta}{\dd t}&=\frac{\mathcal{H}}{R^2}+\frac{\mathcal{I}}{R^3}+\OC{-4},
\label{eq-DD-theta}
\end{align}
\end{subequations} 
where 
\begin{subequations} 
\begin{align}
\mathcal{A}&=2E\left[1+\frac{3}{2}(3\nu-1)\frac{E}{c^2}\right],\\    
\mathcal{B}&=GM\left[1+(7\nu-6)\frac{E}{c^2}\right],\\
\mathcal{C}&=-L^2\left[1+2(3\nu-1)\frac{E}{c^2}\right]+(5\nu-10)\frac{G^2M^2}{c^2}\notag\\
&+\frac{1}{c^2}\frac{4EL\sigma_z}{M},\\
\mathcal{D}&=(8-3\nu)\frac{GML^2}{c^2}+\frac{4G}{c^2}\left[2\frac{s_{1z}s_{2z}}{\mu}-L(\sigma_z+2s_z)\right],\\
\mathcal{H}&=L\left[1+(3\nu-1)\frac{E}{c^2}\right]-\frac{1}{c^2}\frac{2E}{M}\sigma_z,\\
\mathcal{I}&=(2\nu-4)\frac{GML}{c^2}+\frac{4G}{c^2}s_z.
\end{align} 
\end{subequations} 
Since Eq. \eqref{eqs_EC_split} has the same functional form as the GR equations (see Eqs. (2.15) and  (2.16) in Ref. \cite{Damour1985}), the motion can be described through five 1PN parameters $\{a_R,e_R,e_t,e_\theta,K\}$, $a_R$ being the semi-major axis, $e_R,e_t,e_\theta$ three eccentricities, and $K$ the term responsible for the precession in the $x-y$ plane. Therefore, the following  \emph{general Damour-Deruelle  parametrization} holds: 
\begin{subequations} \label{orbit_parametrization_GR}
\begin{align}
\omega (t-t_0)&=u-e_t\sin u +\OC{-4},\
\label{eq:omega}\\
R(t)&=a_R(1-e_R \cos u)+\OC{-4},
\label{eq:R}\\
\theta(t)&=\theta_0+2K\arctan\left[\left(\frac{1+e_\theta}{1-e_\theta}\right)\tan \frac{u}{2}\right]
\nonumber \\
&+\OC{-4},
\label{eq:theta}\\
\omega&=\frac{(-\mathcal{A})^{3/2}}{\mathcal{B}},\\
e_t&=\left[1-\frac{1}{c^2}\frac{(8-3\nu)(G M)}{2a_R}\right]e_R,\\
e_\theta&=\left(1-\frac{\mathcal{A}\mathcal{D}}{2\mathcal{B}L^2}-\frac{\mathcal{A}\mathcal{I}}{\mathcal{B}\mathcal{H}}\right)e_R,\\
K&=1+\frac{1}{c^2}\frac{3G^2M^2}{L_0^2},
\end{align}    
\end{subequations}
where $\omega$ is the mean motion and $u$ the eccentric anomaly. 

The above relations permit to reduce the number of independent parameters to only two, which can be $a_R$ and $e_R$. We note, in particular, that the advantage of having written $K$ in its explicit  PN-expanded form consists in the fact that in this way it is completely determined, as it depends on the 0PN expression of the angular momentum $L_0$.

\subsubsection{Source energy and angular momentum}
\label{sec:source_E_and_J}

In this section, we aim to derive the 1PN-accurate expressions of the source energy $E_{\rm source}$ and orbital angular momentum $L_{\rm source}$, since their time variation enters the balance equations \eqref{balance-equations}. Starting from Eq. \eqref{eqs_EC_split} and performing a conchoidal transformation along the lines of Ref. \cite{Damour1985} (see also Sec. 10.1.5 in Ref. \cite{Poisson-Will2014}, for more details), we obtain the following PN expansions: 
\begin{subequations}\label{eq:source_E_and_J}
\begin{align}
E_{\rm source}&=\mathcal{E}_{\rm 0PN}+\frac{1}{c^2}\left(\mathcal{E}_{\rm 1PN}+\mathcal{E}_{\rm SO}+\mathcal{E}_{\rm SS}\right)\label{eq:orbit_energy_EC} + \OC{-4},\\
L_{\rm source}&=\mathcal{L}_{\rm 0PN}+\frac{1}{c^2}\left(\mathcal{L}_{\rm 1PN}+\mathcal{L}_{\rm SO}+\mathcal{L}_{\rm SS}\right) + \OC{-4},\label{eq:orbit_angular_momentum_EC}
\end{align}
\end{subequations}
where the GR terms are (cf. Eq. (10.20) in Ref. \cite{Poisson-Will2014})
\begin{subequations}
\begin{align}
\mathcal{E}_{\rm 0PN}&=-\frac{G \mu  M}{2 a_R},\\
\mathcal{E}_{\rm 1PN}&=\frac{G^2M^2 \mu  (7-\nu)}{8a_R^2},\\
\mathcal{L}_{\rm 0PN}&=\mu\sqrt{GMa_R \left(1-e_R\right)},\\
\mathcal{L}_{\rm 1PN}&=\frac{G^2 M^2 }{2 \mathcal{L}_{\rm 0PN}}\left[4+(2-\nu)e_R^2\right],
\end{align}
\end{subequations}
while the EC spin-orbit (SO) and spin-spin (SS) contributions are given by
\begin{subequations}
\begin{align}
\mathcal{E}_{\rm SO}&=-\frac{G^2 \mu  M}{\sqrt{G M a_R^5(1-e_R^2)}}(2s_z+\sigma_z),\\
\mathcal{E}_{\rm SS}&=\frac{2 G }{a_R^3 \left(1-e_R^2\right)}s_{1z}s_{2z},\\
\mathcal{L}_{\rm SO}&=-\frac{2 G^2 M }{\mu \mathcal{L}_{\rm 0PN}^2}\left[\left(3+e_R^2\right) s_z+2 \sigma _z\right],\\
\mathcal{L}_{\rm SS}&=\frac{2 G^2 M s_{1z} s_{2z}}{\mu^3\mathcal{L}_{\rm 0PN}^{3}}\left(3+e_R^2\right).
\end{align}
\end{subequations}

\subsubsection{Time average of  energy and angular momentum fluxes}
\label{sec:mean_fluxes}

The last ingredient for constructing the balance equations is represented by the time average of the energy and angular momentum fluxes. At this point, it is necessary to provide a clarification on the strategy to follow. Indeed, in principle there could be two possible routes to pursue. A first approach relies on  expressing $\left<\mathcal{F}\right>$ and $\left<\mathcal{G}\right>$ in terms of $e_R$ and $a_R$ (notice that this is the procedure pioneered by Peters and Mathews in Refs. \cite{Peters1963,Peters1964}). A second scheme consists in writing $\left<\mathcal{F}\right>$ and $\left<\mathcal{G}\right>$ in terms of $E$ and $L$ (in this case an example is furnished by Eq. (4.21) of the Blanchet and Sch{\"a}fer paper \cite{Blanchet-Schafer1989}). We have verified that the two techniques are equivalent in EC theory. However, since we aim to find the evolution equations for $a_R$ and $e_R$, we choose the first pattern, which turns out to be quicker and more straightforward. 

The mediated fluxes of the energy and angular momentum can be written in compact form as
\begin{subequations}\label{eq:time_average_F_and_G}
\begin{align}
-\left<\mathcal{F}\right>&=\frac{1}{c^5}\biggl[\mathscr{F}_{\rm 0PN}+\frac{1}{c^2}\left(\mathscr{F}_{\rm 1PN}+\mathscr{F}_{\rm SO}+\mathscr{F}_{\rm SS}\right) 
 \nonumber \\
&  + \OC{-3}\biggr],
\label{eq:time-average-energy-flux}\\
-\left<\mathcal{G}\right>&=\frac{1}{c^5}\left[\mathscr{G}_{\rm 0PN}+\frac{1}{c^2}\left(\mathscr{G}_{\rm 1PN}+\mathscr{G}_{\rm SO}+\mathscr{G}_{\rm SS}\right) + \OC{-3}\right].
\label{eq:time-average-angular-momentum-flux}
\end{align}
\end{subequations}
In Appendix \ref{app:fluxes}, we report the expressions of $\mathcal{F}$ and $\mathcal{G}$ and delineate the method  to work out $\left<\mathcal{F}\right>$ and $\left<\mathcal{G}\right>$. 

\subsubsection{Differential equations for $a_R(t)$ and $e_R(t)$}
\label{sec:balance_equations}

At this stage, we have all the elements to learn how the two-body orbit shrinks. For this reason, we promote the orbital parameters to functions of time, i.e., $a_R=a_R(t)$ and $e_R=e_R(t)$. Then, starting from Eqs. \eqref{balance-equations}, \eqref{eq:source_E_and_J}, and \eqref{eq:time_average_F_and_G}, we obtain, after a lengthy calculation, 
\begin{subequations}\label{eq:decay_evolution}
\begin{align}
\frac{\dd a_R(t)}{\dd t}&=\frac{1}{c^5}\biggl[\mathcal{P}_{\rm 0PN}+\frac{1}{c^2}\left(\mathcal{P}_{\rm 1PN}+\mathcal{P}_{\rm SO}+\mathcal{P}_{\rm SS}\right)
 \nonumber \\
&+\OC{-3}\biggr],
\label{eq:aR_evolution}\\
\frac{\dd e_R(t)}{\dd t}&=\frac{1}{c^5}\biggl[\mathcal{Q}_{\rm 0PN}+\frac{1}{c^2}\left(\mathcal{Q}_{\rm 1PN}+\mathcal{Q}_{\rm SO}+\mathcal{Q}_{\rm SS}\right)
 \nonumber \\
&+\OC{-3}\biggr],\label{eq:eR_evolution}
\end{align}
\end{subequations}
where the explicit expressions of the above coefficients are given in Appendix \ref{app:balance}.

\subsubsection{Orbit and gravitational waveform}
\label{sec:orbit_and_waveform}

The system \eqref{eq:decay_evolution} is highly non-linear and hence its analytical resolution is demanding. Therefore, we must resort to a numerical routine:  we first consider the ratio between Eqs. \eqref{eq:aR_evolution} and \eqref{eq:eR_evolution}, which leads to a differential equation for the function $a_R(e_R)$; then, upon substituting  the ensuing solution  in Eq. \eqref{eq:eR_evolution}, we determine $e_R(t)$ and hence also $a_R(t)$; finally, this permits to evaluate the relative radius $R(t)$ (cf. Eq. \eqref{eq:R}) and the polar angle $\theta(t)$ (cf. Eq. \eqref{eq:theta}).

During the  inspiral, $e_R(t)$ decreases from its initial value $e_0$ up to zero. This defines the \emph{circularization time} $T_{\rm circ}$, which can be calculated as follows
\begin{equation}
T_{\rm circ}:=\int_0^{e_0}-\frac{\dd t}{\dd e_R} \dd e_R.    
\end{equation}
We can then determine $R(T_{\rm circ})\equiv a(T_{\rm circ})$ and $\theta(T_{\rm circ})$, which will be crucial in the investigation of Sec. \ref{sec:BR_EC_1PN_circular}. 

Having at our disposal all the needed quantities, we can finally compute the two-body orbits via Eq. \eqref{position-vectors-r1-r2-with-spin} and the related gravitational waveform (cf. Eq. \eqref{gravitational_wave_amplitude}) for times $t_0\leq t\leq T_{\rm circ}$. 

\subsection{Quasi-circular orbit}
\label{sec:BR_EC_1PN_circular}

This section is dedicated to quasi-circular orbits. Here, we can distinguish two different alternatives: (1) $e_0\neq0$, then the quasi-elliptic solution must be smoothly connected with the quasi-circular one, which eventually evolves up to the coalescence;  (2)  $e_0=0$, which implies that only the quasi-circular motion has to be taken into account until the bodies collide. Both scenarios entail  a simple mathematical treatment, because we have $a_R=R$ and $e_R=e_t=e_\theta=0$, meaning that  we have to solve only the falling-off differential equation for the radius and then determine the coalescence time. 

The orbits of the two companions and the emitted waveform are  calculated in Sec. \ref{sec:radius_decay}, where we employ Eq. \eqref{eq-DD-theta} to calculate the polar angle $\theta$ via the knowledge of the radius $R(t)$, which is computed numerically. In Sec. \ref{sec:orbital_phase_and_frequency}, we follow a different but equivalent approach to derive the analytical expressions of the orbital phase and frequency. These observable quantities  are extremely useful for the study of the strong relativistic effects imprinted in the GW signal  \cite{Blanchet1996,Blanchet2006b}. 

\subsubsection{Orbit and gravitational waveform}
\label{sec:radius_decay}

The radius evolution is ruled by Eq. \eqref{eq:aR_evolution}, where we set $e_R=0$. This equation is then   integrated numerically to obtain $R(t)$ using $R(T_{\rm circ})$ as initial condition. The polar angle  can be simply evaluated  by setting $\theta(t)=\theta_0+Ku$ (cf. Eq. \eqref{eq:theta}), where $\theta_0=\theta(T_{\rm circ})$. Here, the eccentric anomaly is given by (cf. Eq. \eqref{eq:omega})
\begin{equation}
u=\begin{cases}
\omega(t-T_{\rm circ}),& \mbox{if}\ e_0\neq0,\\
\omega(t-t_0),& \mbox{if}\ e_0=0.
\end{cases}
\end{equation}

The knowledge of $R(t)$ permits to work out the \emph{coalescence time} $T_{\rm coal}$, which is defined as \cite{Peters1964}
\begin{equation}
T_{\rm coal}:=\int_{R_{\rm coal}}^{R(T_{\rm circ})}-\frac{\dd t}{\dd R}\dd R.    
\end{equation}
Here, we set $R_{\rm coal}=2GM/c^2$ in the BH case, since we  suppose that the coalescence occurs when the event horizons have their first contact; on the other hand,  we use $R_{\rm coal}=6GM/c^2$ for NSs, since this is the value  generally accepted as the NS surface.

The orbits and the gravitational waveform of the bodies can now be  easily calculated in the interval $T_{\rm circ}\le t\le T_{\rm coal}$, since the dynamics is fully governed by the function $R(t)$. 

\subsubsection{Orbital frequency and phase}
\label{sec:orbital_phase_and_frequency}

The study of circular orbits in the final stage of compact binaries' evolution is fundamental for inquiring gravity in the strong-field regimes. This implies that high-order relativistic effects should be buried in the GW signal. In order to unearth them, we employ a strategy equivalent to the one developed in Sec. \ref{sec:radius_decay}. The crucial difference with respect to the previous approach relies upon the fact that here we deal with the analytical expressions of the orbital frequency and phase, which are key observables for GW detectors. 

To this end, we define the PN parameters 
\begin{subequations} 
\begin{align}
x &:= \left( \frac{GM\Omega}{c^3} \right)^{2/3},
\label{x-def}
\\
\gamma &: =\frac{GM}{Rc^2},
\end{align}
\end{subequations}    
where $\Omega$ is the source orbital frequency defined as \cite{Maggiore:GWs_Vol1}
\begin{equation} \label{Omega-def}
\int_{t_0}^t \Omega(t') \dd t' =:\theta,
\end{equation}
whose quadratic expression, at 1PN order, reads as 
\begin{align} \label{Omega-squared-PN}
\Omega^2&=\frac{GM}{R^3}\left\{1+\gamma\left[(\nu-3)-\frac{2   \left(2 s_z+3 \sigma _z\right)}{G M^2}c\gamma^{1/2}\right. \right.\notag\\
&\left.\left.+\frac{12  s_{1 z} s_{2 z} }{G^2 M^3 \mu}c^2\gamma \right] +\OC{-4} \right\}.   
\end{align}
Upon inverting the above relation, we obtain the expansion of $\gamma$ in terms of $x$ as follows
\begin{align} \label{x-factor -PN}
    \gamma&=x\left\{ 1 + x \left[\left(1-\frac{\nu }{3}\right) + \frac{2  \left(2 s_z+3 \sigma _z\right) }{3 G
   M^2}c x^{1/2}   \right.\right.\notag\\
   & \left.\left. -\frac{4  s_{1 z} s_{2 z} }{G^2 M^3\mu}c^2 x \right]+ \OC{-4}\right\}.
\end{align}

The source energy \eqref{eq:orbit_energy_EC} can be opportunely rewritten in terms of $\gamma$, and can be then expanded in series of $x$ thanks to Eq. \eqref{x-factor -PN}. With  1PN accuracy, we obtain 
\begin{align} \label{eq:energy_x}
    E_{\rm source}&=-\frac{c^2\mu x}{2} \left\{1-x \left[\frac{(9+\nu)}{12} -\frac{4  \left(4 s_z+3\sigma _z\right)}{3G M^2}c x^{1/2}\right. \right.\notag\\
    &\left. \left. + \frac{8  s_{1z} s_{2z} }{G^2 M^3 \mu }c^2x\right]+\OC{-4}\right\}.
\end{align}
Performing similar calculations also for the energy flux (cf. Eq. \eqref{flux-appendix}), we can rewrite it in terms of $x$ as follows
\begin{align} \label{eq:flux_x}
    \mathcal{F} &=\frac{32 c^5}{5 G}\nu^2 x^5\Biggr{[}\mathbb{F}_{\rm 0PN}+x\Biggr{(}\mathbb{F}_{\rm 1PN}+cx^{1/2}\mathbb{F}_{\rm SO}+c^2x\mathbb{F}_{\rm SS}\Biggr{)}\notag\\
    &+\OC{-3}\Biggr{]},
\end{align}
where 
\begin{subequations}
\begin{align}
\mathbb{F}_{\rm 0PN}&=1\\ 
\mathbb{F}_{\rm 1PN}&= -\left(\frac{1247}{336}+\frac{35 \nu }{12}\right),\\ 
\mathbb{F}_{\rm SO}&=-\frac{11 s_z+5
   \sigma _z}{2 G M^2}, \\
   \mathbb{F}_{\rm SS}&=\frac{\xi_z^2+60 s_{1z} s_{2 z}}{4 G^2 M^3 \mu }.
\end{align}
\end{subequations}

To work out the   variation laws of the orbital phase and frequency for an inspiralling binary system in quasi-circular motion, we resort again to  balance arguments, generalizing the approach pursued in Ref. \cite{Blanchet1996}. Therefore, we first define the adimensional time variable 
\begin{equation}
\vartheta :=\frac{c^3 \nu  t_R}{5 G M},    
\end{equation}
$t_R=t-R/c$ being the retarded time. After that, we exploit the energy balance equation \eqref{E_balance}, where we substitute Eqs. \eqref{eq:energy_x} and \eqref{eq:flux_x} to get $\dd x/\dd\vartheta$. Then, considering $\dd \theta=\Omega \dd t_R=(5/\nu)x^{3/2} \dd \vartheta$, we finally obtain $\dd \theta/\dd x$. These expressions yield two differential equations, which can be exactly integrated in terms of $\Theta:=\vartheta_c-\vartheta$,  $\vartheta_c$ being the (adimensional) coalescence time (recall that $\Theta = \OC{8}$ \cite{Blanchet2014}). Thus, the sought-after analytical formulas of $x(\Theta)$ and $\theta(\Theta)$ at 1PN order in EC theory read as 
\begin{align}
x(\Theta)&=\frac{\Theta^{-1/4}}{4 }\Biggr{[}1-\frac{\mathbb{A}_{\rm 1PN}}{768}\Theta^{-1/4}-\frac{\mathbb{A}_{\rm SO}}{1280}c\Theta ^{-3/8}\notag\\
&-\frac{\mathbb{A}_{\rm SS}}{2048}c^2\Theta^{-1/2}+\OC{-3}\Biggr{]}, 
\label{x-of-Theta}\\
\theta(\Theta) &= -\frac{\Theta^{5/8}}{\nu}\Biggr{[} 1- \frac{5\mathbb{A}_{\rm 1PN}}{1536}\Theta^{-1/4} - \frac{3\mathbb{A}_{\rm SO}}{1024}c\Theta^{-3/8} \notag \\
& - \frac{15 \mathbb{A}_{\rm SS}}{4096}c^2\Theta^{-1/2}+\OC{-3}\Biggr{]},\label{theta-of-Theta}
\end{align} 
where 
\begin{subequations}
\begin{align}
    \mathbb{A}_{\rm 1PN} &=- \frac{4}{21} (743+924 \nu ), \\
    \mathbb{A}_{\rm SO} &= -\frac{32 \left(113 s_z+75 \sigma _z\right)}{3 G M^2},\\
    \mathbb{A}_{\rm SS} &=\frac{16 \left(\xi _z^2+160 s_{1z} s_{2z}\right)}{G^2 M^3\mu}.
\end{align}
\end{subequations}

All the relations displayed in this section formally agree with the corresponding GR ones \cite{Blanchet2006b} up to a normalization factor in the spin. This is something expected in light of the results achieved in our previous studies \cite{Paper3,Paper4,Paper5}. However, it should be noted that the EC spin contributions are multiplied by some powers of $c$, while  in GR this does not occur. The reason of this formal difference is due to the fact that our PN counting of the EC spin differs from the one adopted in GR for the macroscopic angular momentum, as one can promptly verify by consulting Eq. (1.1) in Ref. \cite{Blanchet2006a}. Further details regarding the formal analogy between GR and EC frameworks can be found in Sec. IIIC1 in Ref. \cite{Paper5}.

\subsection{Application:  the event GW150914}
\label{sec:model_application}

\begin{figure*}[bht!]
\hbox{\centering\includegraphics[scale=0.31]{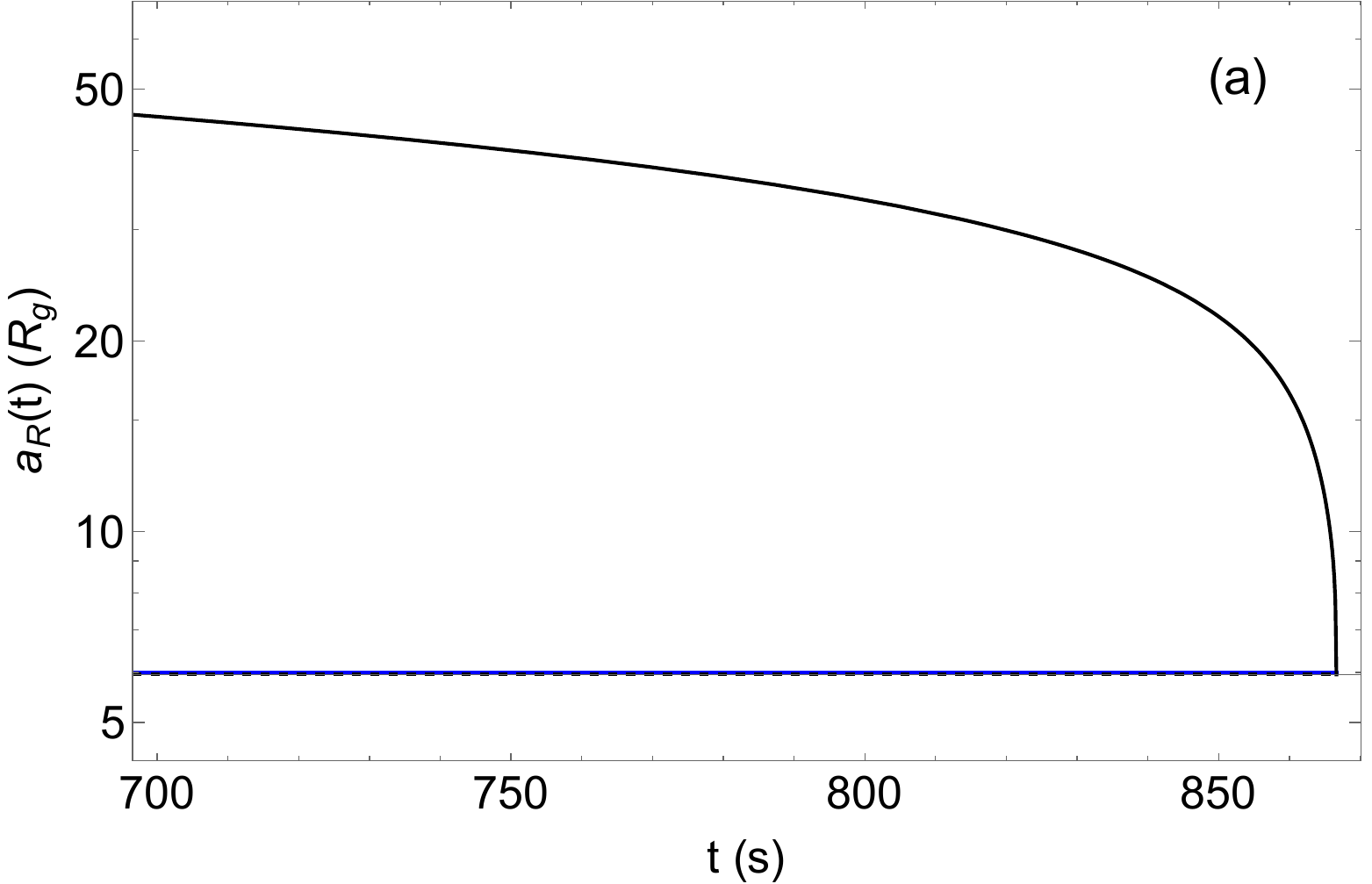}\hspace{1.25cm}
\includegraphics[scale=0.3]{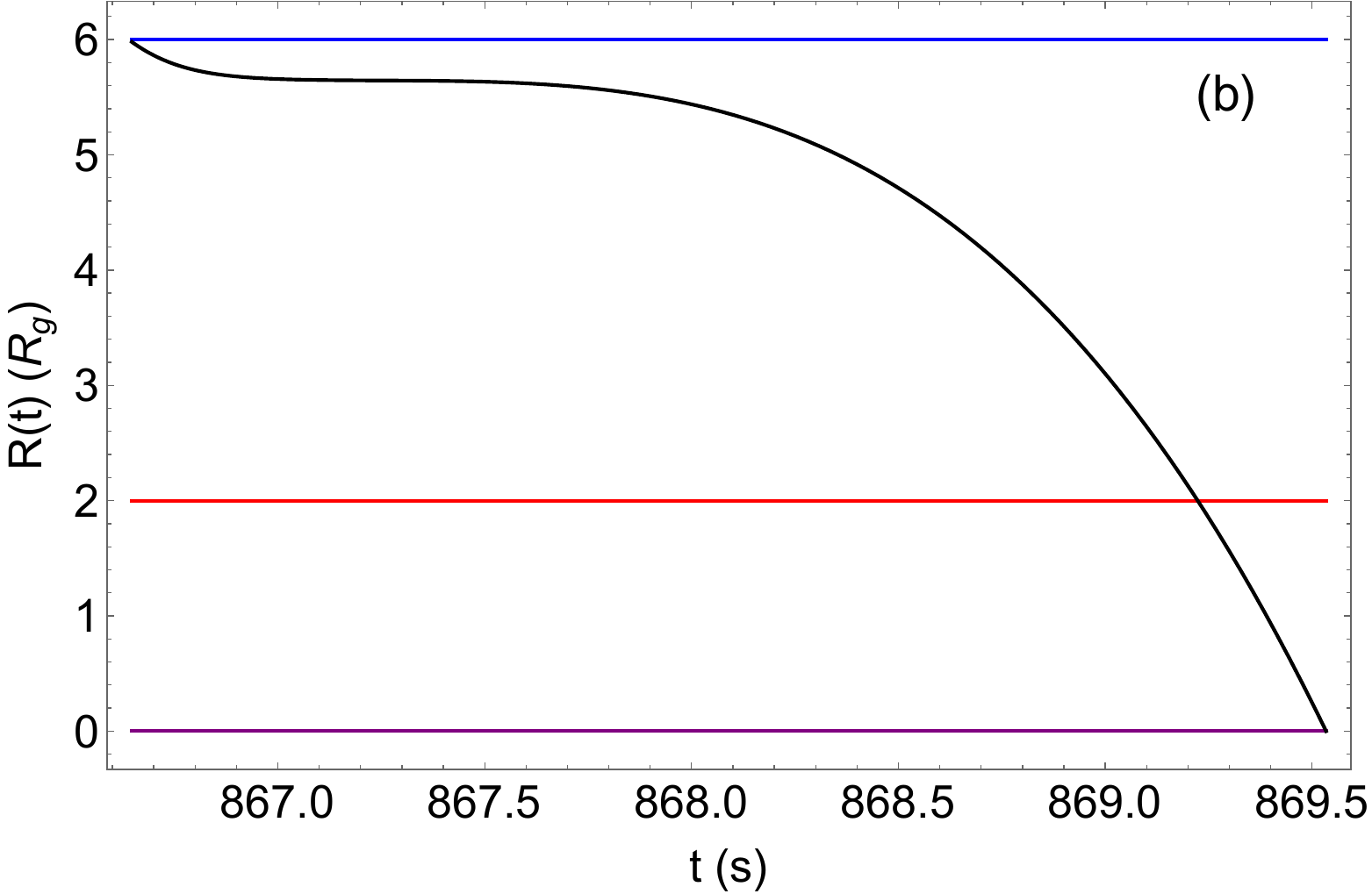}}\vspace{0.3cm}
\hbox{\centering\includegraphics[scale=0.3]{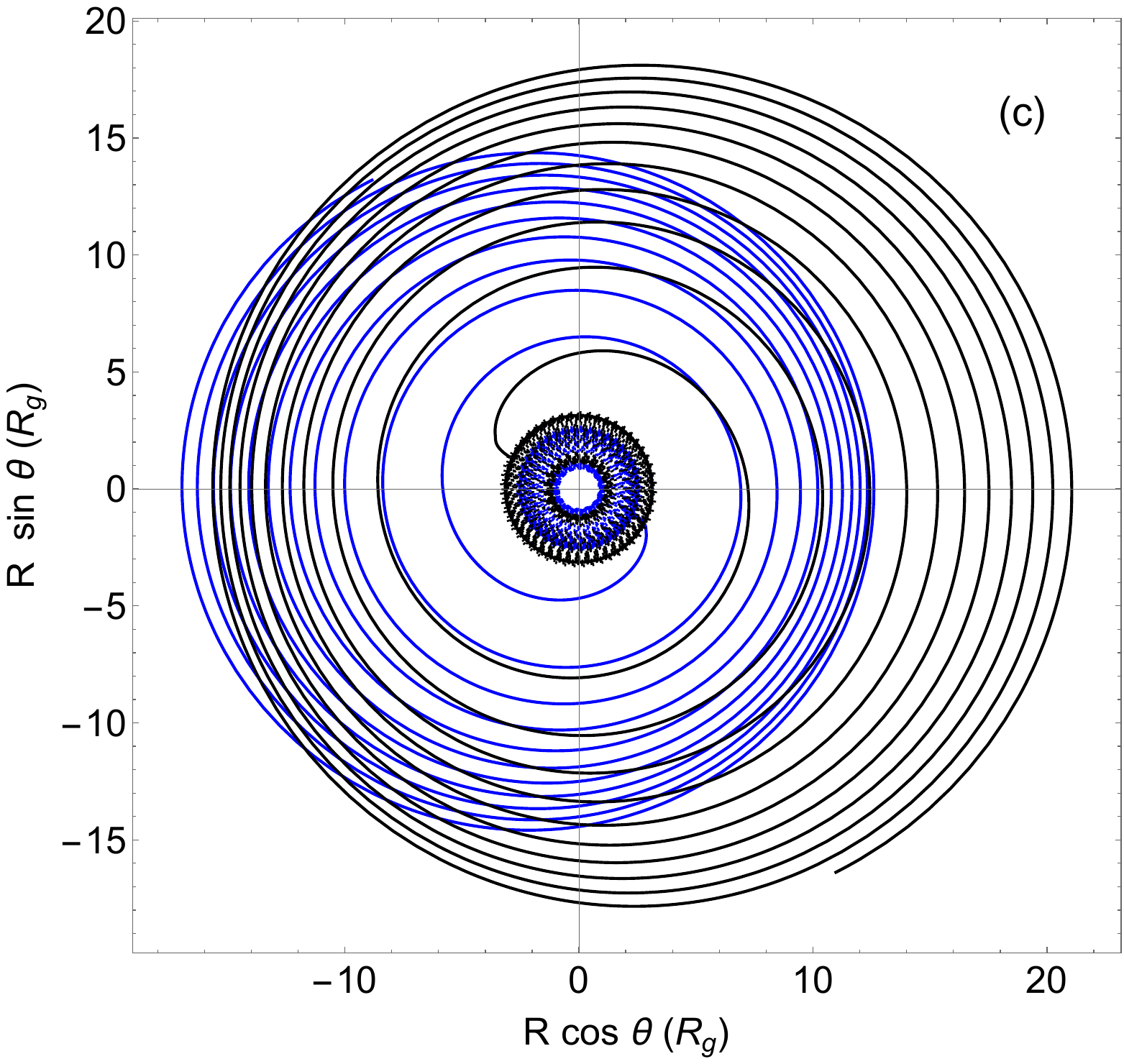}\hspace{0.8cm}
\includegraphics[scale=0.33]{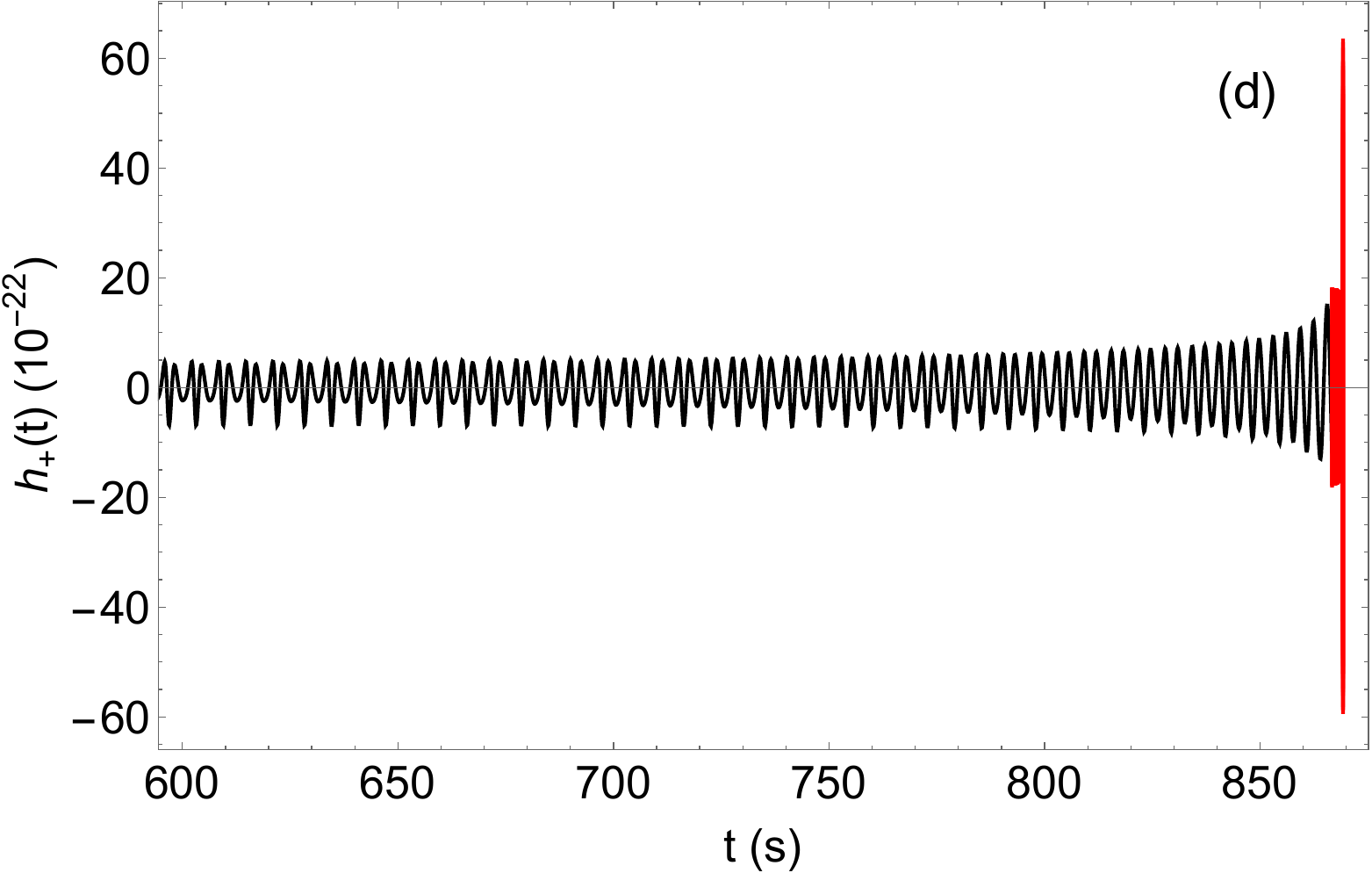}}
\caption{Plots related to the event GW150914, whose parameters can be read off  from Table \ref{tab:Table1}. \emph{Panel (a).} $R$ vs. $t$ for the quasi-elliptic dynamics and with $700$ s $\leq  t \leq T_{\rm circ}$. The blue line is at $R_{\rm ISCO}$ and the dashed one at $a(T_{\rm circ})$. \emph{Panel (b).} $R$ vs. $t$ for quasi-circular orbits, where the blue, red, and purple lines are located  at $R_{\rm ISCO}$,  $R_{\rm coal}$, and  $R=0$, respectively. \emph{Panel (c).}  Orbits of the two objects starting at $t=800$ s, where  continuous and dotted lines indicate  the quasi-elliptic and the quasi-circular case, respectively. The black (blue) line follows  body 1 (2). \emph{Panel (d).} Plus polarization waveform $h_+(t)$ plotted for $600$ s $\leq t \leq T_{\rm coal}$,  the black (red) line referring  to the quasi-elliptic (quasi-circular) motion.}
\label{fig:Fig1}
\end{figure*}

In this section, we apply our back-reaction model to the event GW150914, i.e., the first  GW signal ever observed by LIGO and Virgo produced by the merging of two BHs \cite{Abbott2016}. In Table \ref{tab:Table1}, we list the related parameters along with the input variables \eqref{eq:model_parameters} and output quantities. 
\begin{table}[h!]
    \caption{List of parameters related to the binary BH system of the event GW150914. All the input values have been taken from Ref. \cite{Abbott2016}, apart from the initial separation $a_0$ and eccentricity $e_0$, which have been assigned by us. }
    \centering
    \begin{tabular}{|c|c|c|}
         \hline
          & &\\
        {\bf PARAMETERS} & {\bf UNITS} &\hspace{0.3cm} {\bf VALUE}\hspace{0.3cm} \\
         & &\\
         \hline \hline
        $m_1 $ & $M_\odot$ &  36.00 \\
        $m_2$ & $M_\odot$  & 29.00\\
        $M$ & $M_\odot$  & 65.00\\
        $\mu$ & $M_\odot$  & 16.06 \\
        $\nu$ & & 0.247 \\  
        $s_{\rm 1z}$ & $\hbar$ & $5.10\times10^{59}$\\
        $s_{\rm 2z}$& $\hbar$  & $2.67\times10^{59}$\\
        $d$& ${\rm Mpc}$  &410.00\\
        $R_g$ & m  & $9.64\times10^4$ \\ 
        $\theta_0$ & rad & 0\\
        $t_0$ & s & 0\\
         \hline  \hline
        $a_0$ & $R_g$  &100.00\\
        $e_0$ &  & 0.60\\
        \hline  \hline
        $T_{\rm circ}$ & s & 866.66 \\
        $a(T_{\rm circ})$ & $R_g$ & 5.69\\
        $T_{\rm coal}$ & s & 871.11\\        
        \hline
    \end{tabular}
    \label{tab:Table1}
\end{table}

In Fig. \ref{fig:Fig1}, we present some plots describing the essential features of the  binary BH system dynamics. In \emph{panel (a)}, we display the trend of  $R(t)$ in the quasi-elliptic motion. In this phase, the eccentricity  $e_R(t)$  monotonically decreases (likewise the relative distance $a_R(t)$) and  attains the minimum $e_R=0$ at the time $T_{\rm circ}=866.66$ s, where  the transition to the quasi-circular orbit occurs. In correspondence of this event, we have $R(T_{\rm circ})\equiv a(T_{\rm circ})=5.69\ R_g$ (with $R_g :=GM/c^2$), which is a value close to the innermost stable circular orbit (ISCO) radius $R_{\rm ISCO}=6R_g$, as it is generally confirmed in the literature \cite{Baumgarte2000}. In addition, it is clear that for $t \sim T_{\rm circ}$ there is an abrupt decay of $R(t)$ and a rapid increase of the modulus of $\dd R(t)/\dd t$. This sudden change in the separation and velocity profiles is due to the passage from the inspiral to the merger stage, where the gravitational attraction  becomes more and more intense. In \emph{panel (b)}, we show the  behaviour of $R(t)$ for $t \geq T_{\rm circ}$ and until the BHs collide.   

\begin{figure}[h!]
    \centering
    \includegraphics[trim=0cm 0cm 0cm 0cm,scale=0.36]{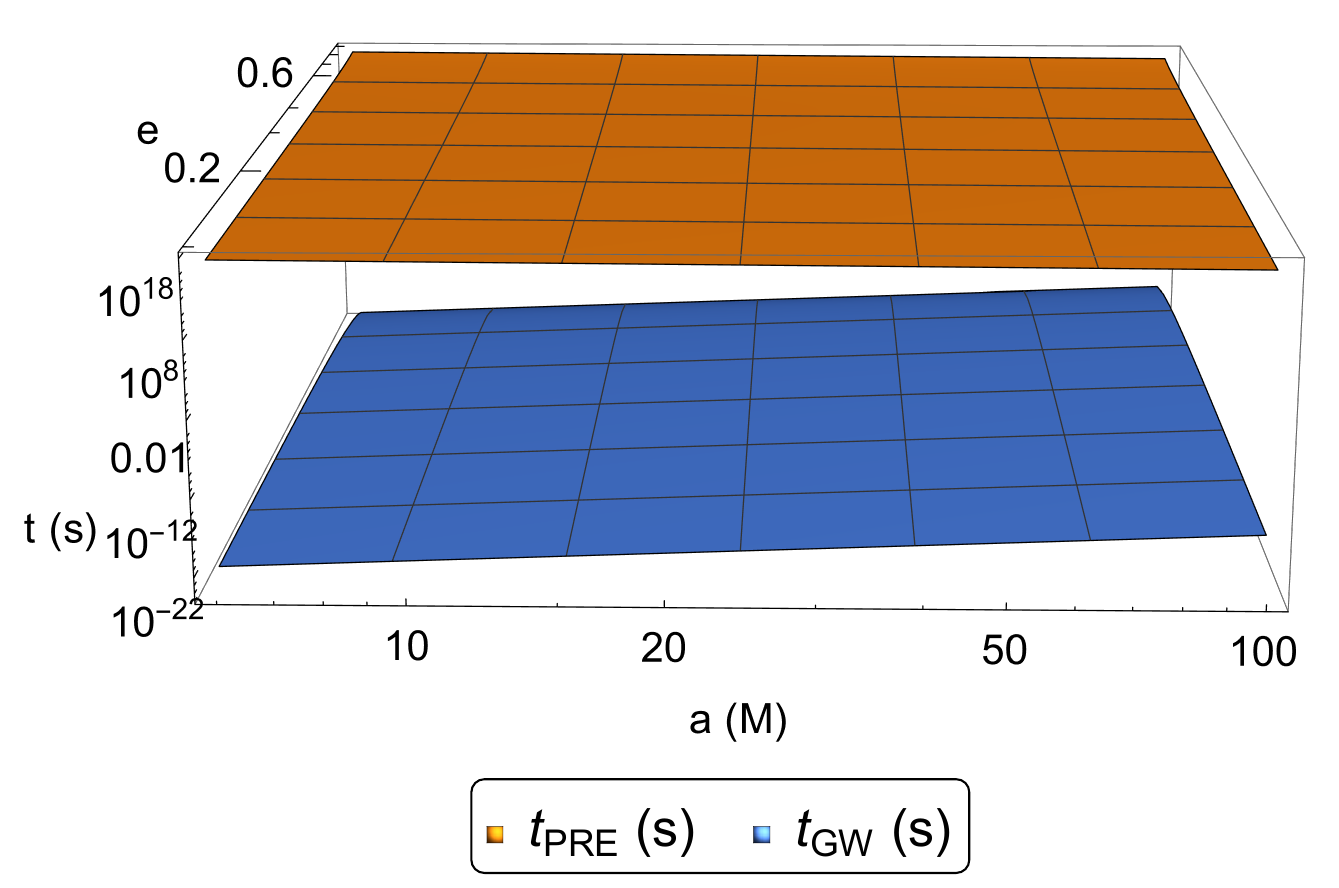}
    \caption{Comparison between the precession and GW 
    emission timescales in terms of the semi-major axis and eccentricity.}
    \label{fig:Fig2}
\end{figure}

The underlying orbits are reported in \emph{panel (c)}. We note that during the quasi-elliptic motion there is no evidence of precession. This is due to the fact that the GW emission occurs on shorter timescales. This statement can be proved by performing the following rough estimation. The shrinkage timescale $t_{\rm GW}$ of the quasi-elliptic orbit is described by Eq. (12) in Ref. \cite{Zwick2020}, whereas the precession timescale $t_{\rm PRE}$ can be calculated by considering the analytical coordinate time $t(\theta)$ \cite{LetteraDBA} evaluated at the periastron shift $\Delta \theta$ (see Eq. (4.15) in Ref. \cite{Damour1985}). The result of our computations can be found in Fig. \ref{fig:Fig2}\footnote{Some clarifications about Fig. \ref{fig:Fig2} are in order. For $t_{\rm PRE}$ we have employed its analytical 1PN expression. On the other hand, $t_{\rm GW}$ has been calculated via the 0PN formula, because we have its analytical form. Indeed, in this case the 1PN corrections predicted by EC theory should be computed numerically and, anyway, do not alter our analysis.}, where it is evident that $t_{\rm GW}\ll t_{\rm PRE}$ for all admissible semi-major axis and eccentricity ranges. 

The  GW signal produced by the inspiralling binary system under consideration is a crucial observable  detected by our interferometers. Therefore, to give an idea of the related gravitational waveform, we provide in \emph{panel (d)} the profile of its \qm{plus} polarization. During the  quasi-elliptic  stage (colored in black), there is a complete agreement with GR, as we have already stressed in our previous studies \cite{Paper2,Paper5}. On the other hand, in the last  instants of the quasi-circular phase (colored in red) there is a fast blowup of the signal. As we have already stressed, this phenomenon can be attributed to the limits of our model, which furnishes a rough description of the merger  as it does not take into account neither the tidal effects nor the internal structure and the high velocities of the bodies. 

\section{Spin contributions at the merger} 
\label{sec:results}

In the previous sections we have discussed  the theoretical aspects of our  back-reaction model, along with its   implications, which have been examined by considering a known astrophysical event. 
In this section, we further exploit our approach to investigate the  EC  modifications to the GW signal in the merger phase,  in order to make a comparison with the GR contributions. As pointed out before, this furnishes a rough estimation of EC corrections, as it is known that the PN approximation breaks down before the coalescence. 

We begin our analysis by first computing the spin effects in the time domain. To this end, let us define the following parameter:
\begin{align} \label{eq:err_time}
\mathcal{E}&:=\underset{t\in [T_{\rm circ},T_{\rm coal}]}{{\rm Mean}}\varrho(t),  
\nonumber \\
\varrho(t) &:= \left|\frac{h_+(t)-\hat{h}_+(t)}{\hat{h}_+(t)}\right|,
\end{align}
where $h_+(t)$ and $\hat{h}_+(t)$  are the plus polarization waveforms framed  in EC theory and GR, respectively (the latter being obtained by setting $s_{Az}=0$). Notice that $\mathcal{E}$ has been computed as the mean of $\varrho(t)$ evaluated for $t\in [T_{\rm circ},T_{\rm coal}]$. However, we have also considered alternative definitions of $\mathcal{E}$ (e.g., maximum for $ t \in [T_{\rm circ},T_{\rm coal}]$, or the value attained at $t=T_{\rm coal}$), and we have verified that all of them  yield similar results.

\begin{figure*}[ht!]
    \centering
    \hbox{\includegraphics[scale=0.33]{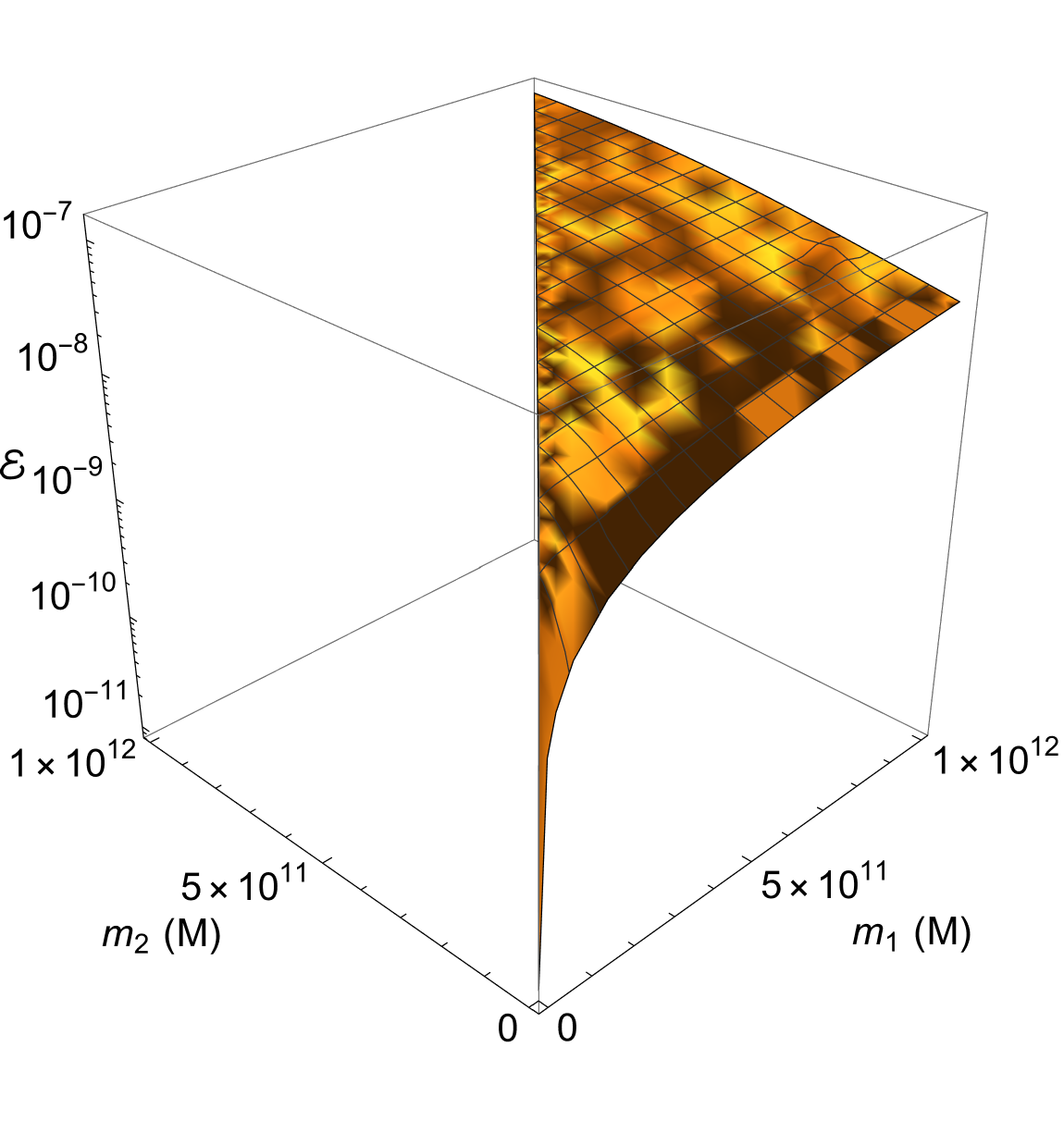}\hspace{0.5cm}
    \includegraphics[scale=0.36]{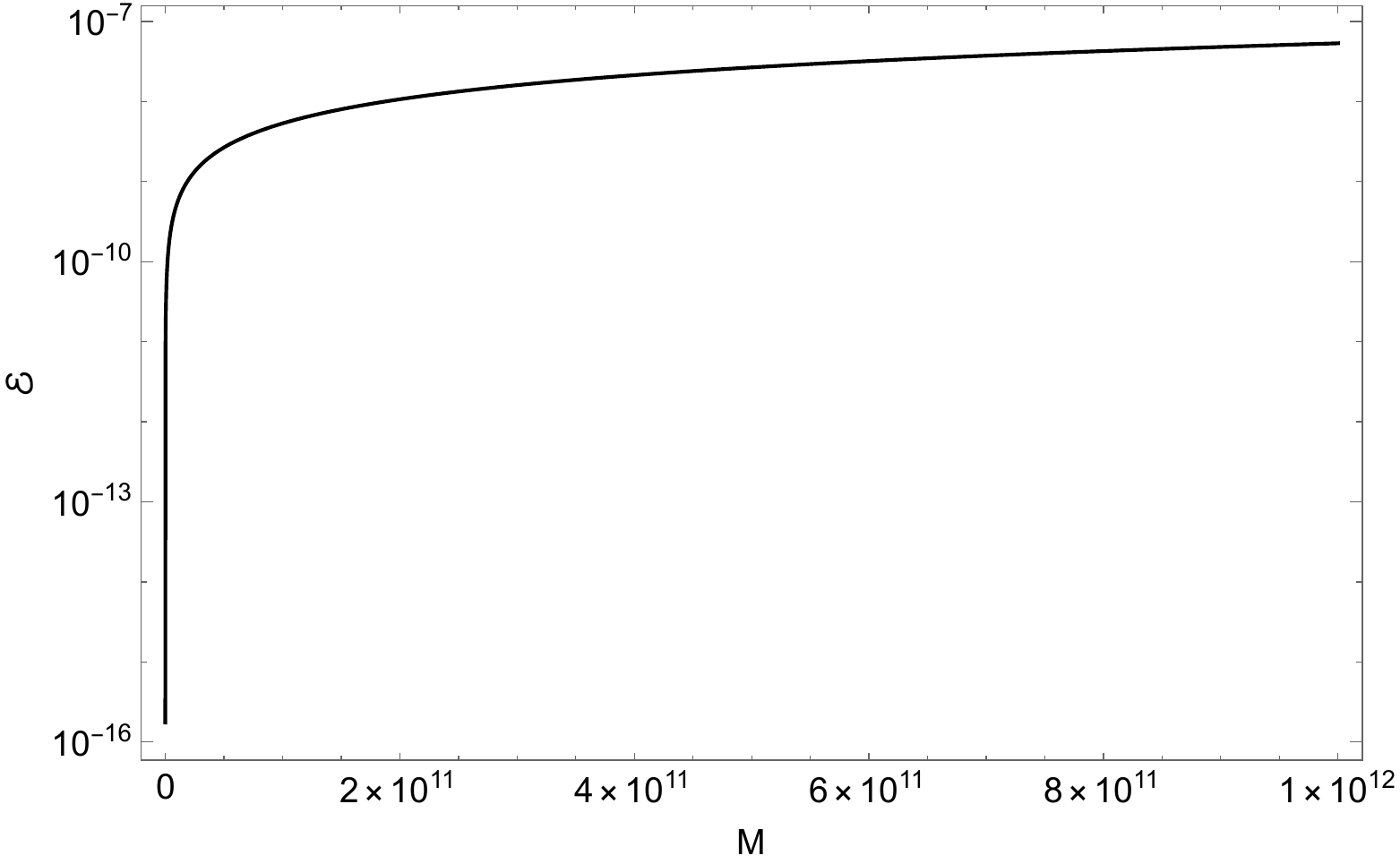}}
    \caption{\emph{Left panel.} Plot of $\mathcal{E}$ in terms of  $m_1$ and $m_2$ (with $m_1\geq m_2$). \emph{Right panel.} $\mathcal{E}$ as a function  of $M$, with $m_1=\frac{2}{3}M$ and $m_2=\frac{1}{3}M$.}
    \label{fig:Fig3}
\end{figure*}

Figure \ref{fig:Fig3} contains both a three-dimensional and two-dimensional plot of $\mathcal{E}$. In the first case, $\mathcal{E}$ is given in terms of $m_1$ and $m_2$, which range in the interval $[3,10^{12}]\ M_\odot$, while in the second we set $m_1=2M/3$ and $m_2=M/3$ and let $M$  vary in  $[18,10^{12}]\ M_\odot$.  From our analysis, we deduce that the EC contributions at the merger are smaller than GR ones by a factor lying between  $10^{-15}$ and $10^{-7}$. It is also clear that the stronger the gravitational field is, the more important the EC corrections become. 

To provide a more comprehensive examination of the EC spin effects, in  Sec. \ref{sec:swa} we perform a  Fourier analysis of the GW signal. 

\subsection{Frequency domain: stationary wave approximation approach}
\label{sec:swa}

In this section, we consider the GW signal in the Fourier frequency space and provide the analytical expression of the phase through the \emph{stationary wave approximation} method  \cite{Damour:1997ub,Damour:2000gg,Damour:2000zb}. This technique is widely used for fitting the GW data and is more reliable in the merger, since it guarantees better convergence properties for the phase.

In this framework, the GW signal can be represented as 
\begin{equation}
    \tilde{h}(f,\boldsymbol{\chi}) = \mathbb{B}(f,\boldsymbol{\chi})e^{-i \phi(f,\boldsymbol{\chi})}, 
\end{equation}
where $\mathbb{B}$ is the amplitude, $f$  the Fourier frequency, and $\boldsymbol{\chi}$ identifies the set of source parameters. 

The phase $\phi$ can be expanded as \cite{Damour:1997ub} 
\begin{align}
    \phi &= 2\pi f t_c -\phi_c -\frac{\pi}{4}+ \frac{3}{128 \nu v^{5}}\Biggr{\{}\phi_{0{\rm{PN}}}\nonumber  \\
    & + \left[\phi_{1{\rm{PN}}} + (cv) \phi_{{\rm{SO}}}+ (cv)^2 \phi_{{\rm{SS}}}\right] v^2+{\rm O}\left(v^{4}\right) \Biggr{\}}, \label{eq:phase_SWA}
\end{align}
where $v=(\pi GMf/c^3)^{1/3}$ is an invariantly defined (adimensional) small velocity, which fulfills the same role as the parameter $x$ (cf. Eq. \eqref{x-def}) in the Fourier domain; in addition, $t_c$ and $\phi_c$ are the time and phase at the coalescence, respectively. The phase coefficients occurring in Eq. \eqref{eq:phase_SWA} read as 
\begin{subequations}
\label{eq:coefficients-phase-stat}
\begin{align}
\phi_{0{\rm{PN}}}&=1, \\
\phi_{1{\rm{PN}}}&=
\frac{3715}{756} + \frac{55}{9}\nu, \\ 
\phi_{{\rm{SO}}}&=\frac{2 \left(113 s_z+75 \sigma _z\right)}{3 G M^2},\\
\phi_{{\rm{SS}}}&=-\frac{5 \left(\xi _z^2+12 s_{1z} s_{2z}\right)}{2 G^2 \mu  M^3}.
\end{align}
\end{subequations}
Notice that also in this case the EC terms $\phi_{{\rm{SO}}}$ and $\phi_{{\rm{SS}}}$ agree formally with those framed in GR up to a multiplicative factor in the spin, and are accompanied by some powers of $c$ because of the  PN counting of the spin adopted in this paper (see comments at the end  of  Sec. \ref{sec:orbital_phase_and_frequency}).

In this context,  the estimation of EC spin corrections can be carried out by defining the following variable:
\begin{equation} \label{eq:err_freq}
\Delta \phi:=\left|\frac{(\phi_{\rm SO}+cv\phi_{\rm SS})cv^3}{\phi_{\rm 0PN}+\phi_{\rm 1PN}v^2}\right|.    
\end{equation}
Substituting in the above formula the coefficients \eqref{eq:coefficients-phase-stat} jointly with the expression \eqref{szi-components-BH-NS} of the spin, and fixing $v=0.6$ (which corresponds to  the typical   velocity achieved by a binary system at the merger \cite{Abbott2016}),  we find, upon setting $M_1=M_2$ with $M_A:=m_A/M_\odot$, 
\begin{equation} \label{eq:EC_estimate}
M_1=10^{20}(2.73 -1.95 \sqrt{1.97 -1.02 \Delta \phi}).
    \end{equation}
Thanks to the above relation, a good compromise between observable masses and reasonable values of $\Delta\phi$ is obtained if $M_1\sim 10^{11}$ and $\Delta\phi\sim10^{-8}$. This implies that the related information in the GW signal can be revealed only by those devices having a frequency sensitivity of the order of $10^{-7}$ Hz \cite{Maggiore:GWs_Vol1}, which unfortunately is beyond the reach of the actual and near-future interferometers. Despite that, it is worth mentioning that the pulsar timing array (PTA) program, which is fulfilling nowadays a crucial role in the   observation of  the stochastic GW background     \cite{EPTA2023,NANOGrav2023,Reardon2023,Bernardo2023a,Bernardo2023b}, is able to attain such low-frequency scales, as it has  a frequency band ranging just below 1 nanohertz to a few tens of microhertz  \cite{Hobbs2010,Tiburzi2018}. However, one should also bear in mind that this kind of observations involve GWs with long wavelengths 
and hence their complete detection would take a time interval roughly between $ 0.32$ and $32$ years.

\section{Conclusions} 
\label{sec:conclusions}

In this work, we have studied the effects of gravitational radiation forces on binary  dynamics   at   1PN order beyond the quadrupole approximation in EC theory, where we have employed  the semiclassical Weyssenhoff fluid for treating the quantum spin effects inside matter. We have adopted a heuristic approach based on the use of balance equations for the energy and the angular momentum, which permit to characterize the secular decaying evolution of PN isolated inspiralling sources through the knowledge of their 1PN equations of motion. This represents our original theoretical contribution within the panorama of extended theories of gravity. It should be noted that in this first investigation we have given a rough estimation of the strong gravitational effects arising during the merger stage, as these should be examined by extending the EOB formalism to EC theory, a task which is out of the scopes of the present article.

Our analysis relies on the back-reaction model devised in Sec. \ref{sec:BR_model}, which works into three steps. First of all, the binary orbit is parametrized \emph{\'a la} Damour-Deruelle in terms of the variables $a_R,e_R$. Then, we derive the expressions of the source energy $E_{\rm source}$ and orbital angular momentum $L_{\rm source}$, as well as the time averages $\langle \mathcal{F}\rangle$ and $\langle \mathcal{G}\rangle$  of the radiation fluxes. Finally, we invoke the balance equations \eqref{balance-equations}, which provide a system of two differential equations for $a_R(t)$ and $ e_R(t)$.

The ensuing investigation can be divided into two scenarios, depending on the value of the initial eccentricity $e_0$. If $e_0\neq0$, the evolution of the bodies begins with a quasi-elliptic orbit, which steadily contracts until it becomes circular, an event occurring at the circularization time $T_{\rm circ}$. Therefore, once the model parameters \eqref{eq:model_parameters} have been assigned, we can numerically integrate Eq. \eqref{eq:decay_evolution} to obtain the $a_R(t)$ and $e_R(t)$ profiles. Then, the motion is smoothly connected to a quasi-circular orbit (having $e_0=0$), which continues until the coalescence time $T_{\rm coal}$. In this case,  we just need to determine $R(t)$  numerically to describe the dynamics. On the other hand, if the initial eccentricity is null, we start directly with a quasi-circular trajectory until the two objects collide. In this second situation, we have also worked out the analytical expressions of the orbital frequency and phase. As an application of our model, we have considered the binary BH merger underlying the event GW150914 (see Fig. \ref{fig:Fig1}). 

In Sec. \ref{sec:results}, we have estimated the EC spin effects featuring the merger phase both in the time and frequency domains. Both analyses have shown that EC corrections are smaller than GR ones and cannot be observed with the actual and near-future GW devices. However, our simple model suggests that EC spin effects could be revealed if we search for supermassive BH binaries (whose masses are of the order of $10^{11}M_\odot$), as they emit nanohertz GWs, which can be potentially detected via the PTA technique.

Despite that, this work is relevant from a theoretical perspective within the panorama of extended models of gravity \cite{Capozziello2011,Clifton2012,Bahamonde2015,Shankaranarayanan2022}. Indeed, to the best of our knowledge, only two examples dealing with GW back-reaction phenomena can be found in the literature, which are framed in: (1) $f(R)$ framework, where the linearized-gravity approach is pursued \cite{DeLaurentis2011,Narang2022}; (2)  scalar-tensor theories,  where  the Blanchet-Damour formalism along with balance arguments are considered \cite{Bernard2022}. Another novel aspect of our paper consists in having setting out a complete description of the binary motion comprising both the quasi-elliptic and the quasi-circular motion, which  have been smoothly connected.

The investigation pursued in this paper has confirmed one of the key aspects of EC theory, namely that
EC spin effects become more important in the strong-field regime characterizing the merger. In this regard, our analysis has revealed that, besides the  masses, the spin density of the bodies also fulfills a  relevant role.  Therefore, this suggests that a more realistic scheme for handling the spin of high-density matter (as it occurs in the NS core) can bring about further observable phenomena. Of course, we should also employ the correct physical model pertaining to the merger to refine our estimations. This entails that the EOB method should be properly extended to EC framework to precisely calculate the spin contributions during the last evolution phases of the binary. This represents a fundamental topic, as confirmed by the recent literature devoted to the applications of the EOB formalism. In particular, in Refs.  \cite{Zhang2020,Shen2023} it is shown that the GW signal during the post-inspiral stages should be correctly interpreted with the appropriate waveform templates in order not to generate fictitious deviations from GR. This demonstrates that the description of the merger phase is an extreme delicate point even in GR. Therefore, the theoretical assessment within extended gravity patterns of the GW phenomena in the strong-field regime is a fascinating challenging task which deserves a careful investigation.

\subsection*{Acknowledgements}
V.D.F., E.B., and S.C. are grateful to Gruppo Nazionale di Fisica Matematica of Istituto Nazionale di Alta Matematica for support. V.D.F., D.U., and S.C. acknowledge the support of INFN {\it sez. di Napoli}, {\it iniziative specifiche} TEONGRAV, MoonLIGHT, and QGSKY. V.D.F. and S.C. acknowledge COST Action CA21136, \textit{Addressing observational tensions in cosmology with systematics and fundamental physics} (CosmoVerse) supported by COST (European Cooperation in Science and Technology). E.B. acknowledges the support of the Austrian Science Fund (FWF) grant P32086. The authors deeply thank Matteo Califano for useful discussions and suggestions. 

\appendix
\section{Energy and angular momentum fluxes and their time average}
\label{app:fluxes}

In this section, we first report the 1PN coefficients of the energy and angular momentum fluxes (cf. Eqs. \eqref{power-radiated-1PN} and \eqref{ang-momentum-flux-1PN});  then, we outline the method to work out the integrals involved in their time averages $\left<\mathcal{F}\right>$ and $\left<\mathcal{G}\right>$ (see Eq. \eqref{eq:time_average_F_and_G}). 

Upon employing the parametrization of the motion illustrated in Sec. \ref{sec:DD_parametrization}, the coefficients of the energy flux 
\begin{align}
\mathcal{F} &=\frac{1}{c^5}\left[\mathcal{F}_{\rm 0PN}+\frac{1}{c^2}\left(\mathcal{F}_{\rm 1PN}+\mathcal{F}_{\rm SO}+\mathcal{F}_{\rm SS}\right) + \OC{-3}\right],
\end{align}
read as
\begin{widetext}
\begin{subequations}
\label{flux-appendix}
\begin{align}
\mathcal{F}_{\rm 0PN}&=-\frac{8 G^2 M\mathcal{E}_{\rm 0PN}^2  }{15 \pi  a_R^3 (1-e_R \cos u)^5}\Biggr{[}24-\Biggr{(}23+\cos (2 u)\Biggr{)}e_R^2\Biggr{]},\\
\mathcal{F}_{\rm 1PN}&=-\frac{\mathcal{E}_{\rm 0PN}^2 G^3 M^2}{105 \pi  a_R^4 (1-e_R \cos u)^7}\Biggr{\{} e_R^4 \Biggr{[}(9046-5131 \nu) +8 (294 \nu +101) \cos (2 u)+(43 \nu +6) \cos (4 u)\Biggr{]}  \notag \\
&+4 e_R^3 \cos (u) \Biggr{[}(637 \nu-7336) +(27 \nu -88) \cos (2 u)\Biggr{]}+8 e_R^2 \Biggr{[}(581 \nu+1815) +(448-429 \nu ) \cos (2 u)\Biggr{]}\notag\\
&+e_R \cos (u)\Biggr{[}32 (17 \nu +420)\Biggr{]}-4 (420 \nu +2927)  \Biggr{\}},\\
\mathcal{F}_{\rm SO}&=-\frac{8G^2 \mathcal{E}_{\rm 0PN}^3}{15 \pi \mathcal{L}_{\rm 0PN} a_R^3 \sqrt{e_R+1} (1-e_R \cos u)^7} \Biggr{\{}e_R^4 \Biggr{[}2 s_z (68 \cos (2 u)+\cos (4 u)+175)+(16 \cos (2 u)+\cos (4 u)+423) \sigma _z\Biggr{]}\notag\\
&-16 e_R^3 \cos (u) \Biggr{[}2 s_z (\cos (2 u)+20)+(\cos (2 u)+13) \sigma _z\Biggr{]}+4 e_R^2 \Biggr{[}s_z (68 \cos (2 u)+48)+(47 \cos (2 u)-115) \sigma _z\Biggr{]}\nonumber\\
&-32 e_R \cos (u) \left(27 s_z+17 \sigma _z\right)+8 \left(73 s_z+75 \sigma _z\right)\Biggr{\}} ,\\
\mathcal{F}_{\rm SS}&= 
-\frac{4 E_0 G^3 Ms_{1z} s_{2z}}{15 \pi  a_R^6 \left(1-e_R^2\right) (1-e_R \cos u)^7}\Biggr{\{}-e_R^4\Biggr{[}671 -24 \cos (2 u)+\cos (4 u)+\frac{6\xi _z^2}{s_{1z} s_{2z}} (2 \cos (2 u)-1) \Biggr{]}\notag\\
&+12 e_R^2 \Biggr{[}79-19 \cos (2 u)+\frac{\xi _z^2}{s_{1z} s_{2z}}\cos (2 u)\Biggr{]}+8e_R^3\Biggr{[} 23 \cos (u)+\cos (3 u)\Biggr{]}+576e_R \cos (u)\notag\\
&-6 \left[140 +\frac{\xi _z^2}{s_{1z} s_{2z}}\right]\Biggr{\}},
\end{align}
\end{subequations}
\end{widetext}
whereas those of the angular momentum flux 
\begin{align}
\mathcal{G}&=\frac{1}{c^5}\left[\mathcal{G}_{\rm 0PN}+\frac{1}{c^2}\left(\mathcal{G}_{\rm 1PN}+\mathcal{G}_{\rm SO}+\mathcal{G}_{\rm SS}\right)+ \OC{-3}\right],
\end{align}\\
are
\begin{widetext}
\begin{subequations}
\begin{align}
\mathcal{G}_{\rm 0PN}&=\frac{4 G^2M\mathcal{E}_{\rm 0PN} \mathcal{L}_{\rm 0PN} \sqrt{1+e_R} }{5 \pi  a_R^{5/2}(1-e_R \cos u)^4}\Biggr{[}e_R^2(\cos (2 u)-5)-4e_R \cos u+8 \Biggr{]},\\
\mathcal{G}_{\rm 1PN}&=-\frac{8 G^2 M \mathcal{E}_{\rm 0PN}^3}{315 \pi  \mathcal{L}_{\rm 0PN} a_R \sqrt{1+e_R} (1-e_R \cos u)^6}\Biggr{\{}-3 e_R^6 \Biggr{[}15 (72 \nu -19)-12 (73 \nu +105) \cos ^2(u)+(92 \nu +314) \cos ^4(u)\Biggr{]}\nonumber\\
&-4 e_R^5 \cos (u) \Biggr{[}(3082+89 \nu) +6 (7 \nu -179) \cos (2 u)\Biggr{]}+e_R^4\Biggr{[}(10279+7280 \nu)-54 (83 \nu +333) \cos ^2(u)\nonumber\\
&+6 (67 \nu +31) \cos ^4(u)\Biggr{]}-4 e_R^3 \cos (u)\Biggr{[}-(7325+214) \nu +3 (7 \nu +232) \cos (2 u)\Biggr{]}+e_R^2\Biggr{[}-(14354+4261 \nu)\nonumber\\ 
&+9 (145 \nu +537) \cos (2 u)\Biggr{]}-e_R\Biggr{[} \cos (u)4 (314 \nu +3109)\Biggr{]}+(1652 \nu +7297)\Biggr{\}} , \\
\mathcal{G}_{\rm SO}& = \frac{2G^2 \mathcal{E}_{\rm 0PN}^2}{15 \pi a_R^3 \left(1-e_R^2\right) (1-e_R \cos u)^6}\Biggr{\{}
e_R^6 \Biggr{[}72 s_z \cos ^2(u)(\cos (2 u)-7)+5\sigma _z (-8 \cos (2 u)+\cos (4 u)-57)\Biggr{]}\nonumber\\
&+e_R^5 \Biggr{[}2 s_z (416 \cos (u)-84 \cos (3 u))+8\sigma _z (15 \cos (u)-8 \cos (3 u))\Biggr{]}+e_R^4 \Biggr{[}2 s_z (144 \cos (2 u)+3 \cos (4 u)-239)\nonumber\\
&+\sigma _z(-44 \cos (2 u)+7 \cos (4 u)+477)\Biggr{]}+e_R^3 \Biggr{[}16 (41 \cos (u)-2 \cos (3 u)) \sigma _z-8 s_z (7 \cos (u)+3 \cos (3 u))\Biggr{]}\nonumber\\
&+e_R^2 \Biggr{[}8 s_z (75 \cos (2 u)+73)+\sigma _z(420 \cos (2 u)-692)\Biggr{]}-e_R \cos (u)\Biggr{[}2120 s_z+1448\sigma _z\Biggr{]}+(968 s_z+920 \sigma _z)\Biggr{\}}, \\
\mathcal{G}_{\rm SS}&=\frac{2G^{5/2} M^{1/2} \mathcal{E}_{\rm 0PN}s_{1z} s_{2z} }{5 \pi  a_R^{9/2} \left(1-e_R^2\right)^{3/2} (1-e_R \cos u)^6} \Biggr{\{}-4 e_R^5 \cos (u) \left[24 \cos ^2(u)+\frac{\xi _z^2}{s_{1z} s_{2z}}\right]+e_R^4 \Biggr{[}507-140 \cos (2 u)+\cos (4 u)\nonumber\\
&+\frac{4\xi _z^2}{s_{1z} s_{2z}}\Biggr{]}+8 e_R^3 \Biggr{[}57 \cos (u)-\cos (3 u)+\frac{\xi _z^2}{s_{1z} s_{2z}}\cos (u)\Biggr{]}-4 e_R^2 \Biggr{[}3 (53-17 \cos (2 u))+\frac{2 \xi _z^2}{s_{1z} s_{2z}}\Biggr{]}\nonumber\\
&-4 e_R \cos (u) \left[152+\frac{\xi _z^2}{s_{1z} s_{2z}}\right]+3 e_R^6 \Biggr{[}-65+16 \cos (2 u)+\cos (4 u)\Biggr{]}+4 \left[116 +\frac{\xi _z^2}{s_{1z} s_{2z}}\right]\Biggr{\}}.
\end{align}
\end{subequations}
\end{widetext}
To obtain the related time-averaged expressions, we should perform an integral in terms of the variable $u$ (i.e, the eccentric anomaly). To compactify the ensuing formulas, we write all terms to be integrated through the following general function: 
\begin{equation}
f(u,l,m,n):=\frac{(-1)^{l+1}\cos(n u)\cos^m u}{(1-e_R\cos u)^{l+1}},    
\end{equation}
where $l,m,n$ are positive integers ranging in the following intervals: $l\in[3,6],m\in[0,2],n\in[0,4]$. Therefore, if we integrate the function $f$, we have for $n=m=0$ \cite{Blanchet-Schafer1989}
\begin{align}
\int_0^{2\pi} f(u,l,0,0) \dd u&=\frac{(-1)^{l+1}}{(1-e_R^2)^{\frac{l+1}{2}}}P_l\left(\frac{1}{\sqrt{1-e_R^2}}\right),
\end{align}
where $P_l(y)$ is the Legendre polynomial of order $l$ \cite{Abramowitz-Stegun(1964)}. In general, after the substitution $y=\cos u$, we
have
\begin{align}
&\int_0^{2\pi} f(u,l,m,n) \dd u= 2 (-1)^{l+1}  \int_{-1}^1 \frac{T_n(y)y^m\ \dd y}{\sqrt{1-y^2}(1-e_R y)^{l+1}}\notag\\
&=2 (-1)^{l+1}\sum_{j=0}^n a_j\int_{-1}^1 \frac{y^{j+m}\ \dd y}{\sqrt{1-y^2}(1-e_Ry)^{l+1}},
\end{align}
where $T_n(y)=\sum \limits_{j=0}^n a_j y^j$ are the Chebyshev polynomials of the first kind, where the coefficients $a_j$ are known once the value of $n$ is fixed \cite{Abramowitz-Stegun(1964)}. Therefore, all the calculations to be performed  reduce to   the  evaluation of the following integral: 
\begin{widetext}
\begin{align}
\int_{-1}^1 \frac{y^k\ \dd y}{\sqrt{1-y^2}(1-e_R y)^{l+1}}&=\frac{\pi}{4} \Biggr{[}2 \Biggr{(}(-1)^k+1\Biggr{)} \Gamma \left(\frac{k+1}{2}\right){}_3\tilde{F}_2\left(\frac{l+1}{2},\frac{2+l}{2},\frac{k+1}{2};\frac{1}{2},\frac{k+2}{2};e_R^2\right)\notag\\
&-e_R (l+1) \Biggr{(}(-1)^k-1\Biggr{)} \Gamma \left(\frac{k}{2}+1\right) \, {}_3\tilde{F}_2\left(\frac{l+2}{2},\frac{l+3}{2},\frac{k+2}{2};\frac{3}{2},\frac{k+3}{2};e^2_R\right)\Biggr{]},
\end{align}
\end{widetext}
 $\Gamma(y)$ being the Euler gamma function and ${}_p\tilde{F}_q\left(\boldsymbol{c};\boldsymbol{d};y\right)$  the regularized generalized hypergeometric function, where $\boldsymbol{c}=\{c_1,\dots,c_p\}\in\mathbb{R}^p$, and $\boldsymbol{d}=\{d_1,\dots,d_q\}\in\mathbb{R}^q$ with $p,q\in\mathbb{N}$ \cite{Abramowitz-Stegun(1964)}.

\section{Balance equations}
\label{app:balance}
In this section, we write explicitly the coefficients occurring in  the balance equation \eqref{eq:decay_evolution}. Those of Eq. \eqref{eq:aR_evolution} read as
\begin{widetext}
\begin{subequations}
\begin{align}
\mathcal{P}_{\rm 0PN}&=\frac{4 G^2 M \mathcal{E}_0  }{15 a_R^2 \left(1-e_R^2\right)^{7/2}}\left(96+292 e_R^2+37 e_R^4\right),\\
\mathcal{P}_{\rm 1PN}&=-\frac{G^3 M^2 \mathcal{E}_0}{210 a_R^3 \left(1-e_R^2\right)^{9/2}} \Biggr{[}(1036 \nu -5501) e_R^6+126 (166 \nu +275) e_R^4+88 (490 \nu +1821) e_R^2+16(1751 + 588\nu)\Biggr{]},\\
\mathcal{P}_{\rm SO}&=\frac{2 G^3M \mathcal{L}_0  \sqrt{1+e_R}}{15 a_R^5 \left(1-e_R^2\right)^{11/2}} \Biggr{\{}\Biggr{[}215 e_R^6+1758 e_R^4+6784 e_R^2+2128\Biggr{]} s_z+\Biggr{[}217 e_R^6+3348 e_R^4+7348 e_R^2+1680\Biggr{]} \sigma _z\Biggr{\}},\\
\mathcal{P}_{\rm SS}&=-\frac{G^3 Ms_{1z} s_{2z}}{15 a_R^5 \left(1-e_R^2\right)^{11/2}} \Biggr{\{}e_R^6 \Biggr{[}1684+\frac{27 \xi _z^2}{s_{1z} s_{2z}}\Biggr{]}+18 e_R^4 \Biggr{[}1248+\frac{23 \xi _z^2}{s_{1z} s_{2z}}\Biggr{]}+8 e_R^2 \Biggr{[}5300+\frac{57 \xi _z^2}{s_{1z} s_{2z}}\Biggr{]}\notag\\
&+48 \Biggr{[}188+\frac{\xi _z^2}{s_{1z} s_{2z}}\Biggr{]}\Biggr{\}},
\end{align}
\end{subequations}
\end{widetext}
while for Eq. \eqref{eq:eR_evolution} we have
\begin{widetext}
\begin{subequations}
\begin{align}
\mathcal{Q}_{\rm 0PN}&=\frac{2G^2 M\mathcal{E}_0  }{15 a_R^3 e_R\left(1-e_R^2\right)^{5/2}}\left(121 e_R^2+304\right),\\
\mathcal{Q}_{\rm 1PN}&= -\frac{G^3 M^2\mathcal{E}_0 }{1260 a_R^4 e_R
   \left(1-e_R^2\right)^{7/2}} \Biggr{[}(9492 \nu -75465) e_R^6+12 (8029 \nu +27344) e_R^4+8 (14476 \nu +50003) e_R^2+448 (4 \nu -1)\Biggr{]},\\
\mathcal{Q}_{\rm SO}&=-\frac{2 \mathcal{E}_0 G^{5/2}M^{1/2}}{15 a_R^{9/2} e_R \left(e_R^2-1\right)^4}\Biggr{\{}\Biggr{[}766 e_R^6+3351 e_R^4+6760 e_R^2+8\Biggr{]} s_z+\Biggr{[}828 e_R^6+5733 e_R^4+6040 e_R^2-8\Biggr{]} \sigma _z\Biggr{\}},\\
\mathcal{Q}_{\rm SS}&=-\frac{G^3 M  s_{1 z} s_{2 z}}{30a_R^6 \left(1-e_R^2\right)^{9/2}}e_R\Biggr{\{}e_R^4 \left[4892+\frac{45 \xi _z^2}{s_{1z} s_{2z}}\right]+12 e_R^2 \left[3002+\frac{45 \xi _z^2}{s_{1z} s_{2z}}\right]+ 24\left[1444 +\frac{15 \xi _z^2}{s_{1z} s_{2z}}\right] \Biggr{\}}.
\end{align}
\end{subequations}
\end{widetext}

\bibliographystyle{spphys}       
\bibliography{references}

\end{document}